\documentclass[10pt,numbers,preprint,nocopyrightspace]{sigplanconf}
\usepackage{epsfig}
\usepackage{times}
\usepackage[dvipsnames]{xcolor}
\usepackage{xspace}
\usepackage{boxedminipage}
\usepackage{pseudocode}
\usepackage{multirow}
\usepackage{float}
\usepackage{tabularx}
\usepackage{enumitem}
\usepackage{microtype}
\usepackage[labelfont=bf,skip=2pt,belowskip=2pt]{caption}
\usepackage{subcaption}
\usepackage{algorithm}
\usepackage[noend]{algpseudocode}
\usepackage{url}

\usepackage[compact]{titlesec}
\usepackage[T1]{fontenc}
\usepackage[scaled]{beramono}
\usepackage{listings}

\usepackage{balance}

\usepackage[subtle,margins=normal]{savetrees}

\usepackage{hyperref}

\usepackage{etoolbox}
\makeatletter
\patchcmd{\ttlh@hang}{\parindent\z@}{\parindent\z@\leavevmode}{}{}
\patchcmd{\ttlh@hang}{\noindent}{}{}{}
\makeatother

\conferenceinfo{}{}
\copyrightyear{2017}

\newcommand{\sn}{Weld\xspace}

\newcommand{\il}{IR\xspace}
\newcommand{\ils}{IRs\xspace}

\newcommand{\ie}{\emph{i.e.},\xspace}
\newcommand{\eg}{\emph{e.g.},\xspace}

\newcommand{\etc}{{\it etc.}\xspace}
\newcommand{\eat}[1]{}

\newcommand{\squeezeup}{\vspace{-5mm}}

\definecolor{darkgreen}{rgb}{0.0, 0.5, 0.0}

\newcommand\code[1]{\lstinline$#1$}

\lstdefinelanguage{Python}{
 keywords={typeof, null, catch, switch, in, int, str, float, self},
 keywordstyle=\color{NavyBlue}\bfseries,
 ndkeywords={boolean, throw, import},
 ndkeywords={return, class, if ,elif, endif, while, do, else, True, False , except, def, lambda, print},
 ndkeywordstyle=\color{BrickRed}\bfseries,
 basicstyle=\footnotesize\ttfamily,
 identifierstyle=\color{black},
 sensitive=false,
 comment=[l]{\#},
 morecomment=[s]{/*}{*/},
 commentstyle=\color{ForestGreen}\ttfamily,
 string=[s]{"}{"},
 showstringspaces=false,
 stringstyle=\color{violet}\ttfamily,
}

\lstdefinelanguage{Scala}{
 keywords={typeof, null, catch, switch, in, int, str, float, self},
 keywordstyle=\color{BrickRed}\bfseries,
 ndkeywords={boolean, throw, import},
 ndkeywords={val, return, class, if ,case, endif, while, do, else, True, False , except, def, lambda},
 ndkeywordstyle=\color{NavyBlue}\bfseries,
 basicstyle=\footnotesize\ttfamily,
 identifierstyle=\color{black},
 sensitive=false,
 comment=[l]{\#},
 morecomment=[s]{/*}{*/},
 commentstyle=\color{ForestGreen}\ttfamily,
 stringstyle=\color{red}\ttfamily,
}

\lstdefinelanguage{NVL}{
 keywords={char, int, float, long, double, merger, vecBuilder, vec, dict, dictBuilder, dictMerger, vecMerger, groupBuilder},
 keywordstyle=\color{NavyBlue}\bfseries,
 ndkeywords={for, result, merge, if, else, bitselect, while, lookup, len, sort, toVec, map, flatmap, reduce, filter, groupby, zip, mapAndReduce, call},
 ndkeywordstyle=\color{BrickRed}\bfseries,
 basicstyle=\footnotesize\ttfamily,
 identifierstyle=\color{black},
 sensitive=false,
 comment=[l]{\/\/},
 morecomment=[s]{/*}{*/},
 commentstyle=\color{ForestGreen}\ttfamily,
 stringstyle=\color{red}\ttfamily,
 breaklines=true,
}

\newenvironment{denseenum}{
\begin{enumerate}[topsep=2pt, partopsep=0pt, leftmargin=1.5em]
  \setlength{\itemsep}{4pt}
  \setlength{\parskip}{0pt}
  \setlength{\parsep}{0pt}
}{\end{enumerate}}

\begin{document}

\authorinfo{Shoumik Palkar, James Thomas, Deepak Narayanan, Anil Shanbhag$^{\dagger}$,
Rahul Palamuttam, Holger Pirk$^{\ddagger}$, Malte Schwarzkopf$^{\dagger}$, Saman Amarasinghe$^{\dagger}$, Samuel Madden$^{\dagger}$, Matei Zaharia}
{\emph{Stanford University, MIT CSAIL$^{\dagger}$, Imperial College London$^{\ddagger}$}}

\title{\sn: Rethinking the Interface Between Data-Intensive Libraries}
\maketitle

\subsection*{Abstract}

Data analytics applications combine multiple functions from different libraries and frameworks.
Even when each function is optimized in isolation, the performance of the combined application
can be an order of magnitude below hardware limits due to extensive data movement across these
functions.
To address this problem, we propose \sn, a new interface between data-intensive libraries that
can optimize across disjoint libraries and functions.
\sn exposes a lazily-evaluated API where diverse functions can submit their computations in a
simple but general intermediate representation that captures their data-parallel structure.
It then optimizes data movement across these functions and emits efficient code
for diverse hardware.
\sn can be integrated into existing frameworks such as Spark, TensorFlow, Pandas and NumPy without changing their user-facing APIs.
We demonstrate that \sn can speed up applications using these frameworks by up to 29$\times$.

\lstset{language=NVL}

\section{Introduction}
\label{sec:introduction}

The main way users are productive writing software is by \emph{combining} libraries and functions
written by other developers. This is especially true in data analytics applications, which often need to compose
many disparate algorithms into one workflow. For instance, a typical machine learning pipeline
might select some data using Spark SQL~\cite{sparksql}, transform it using NumPy and Pandas
\cite{numpy,pandas}, and train a model with TensorFlow~\cite{tensorflow}, taking advantage of
Python's rich ecosystem of data science libraries.

Traditionally, the interface for composing these libraries has been function calls. Each library
function takes pointers to in-memory data, performs a computation, and writes a result back to
memory. Unfortunately, this interface is increasingly inefficient for \emph{data-intensive}
applications.  The gap between memory bandwidth and processing speeds has grown steadily over
time~\cite{kagi1996memory}, so that, on modern hardware, many applications spend most of their time
on \emph{data movement} between functions.
For example, even though NumPy and Pandas use optimized C functions (\eg BLAS~\cite{blas}) for their
operators, we find that programs that use multiple such operators can be 8$\times$ slower than
handwritten code, because the function call interface requires materializing intermediate
results in memory after each operation.
This problem gets worse when some libraries use hardware accelerators, such as GPUs, because data movement
into these accelerators is even slower~\cite{semcache}.
In short, the core
interface developers have used to compose software for the past 50 years---functions that exchange
data through memory---misuses the most precious resources on modern hardware.

\begin{figure}[t!]
  \centering
  \includegraphics[width=0.9\columnwidth]{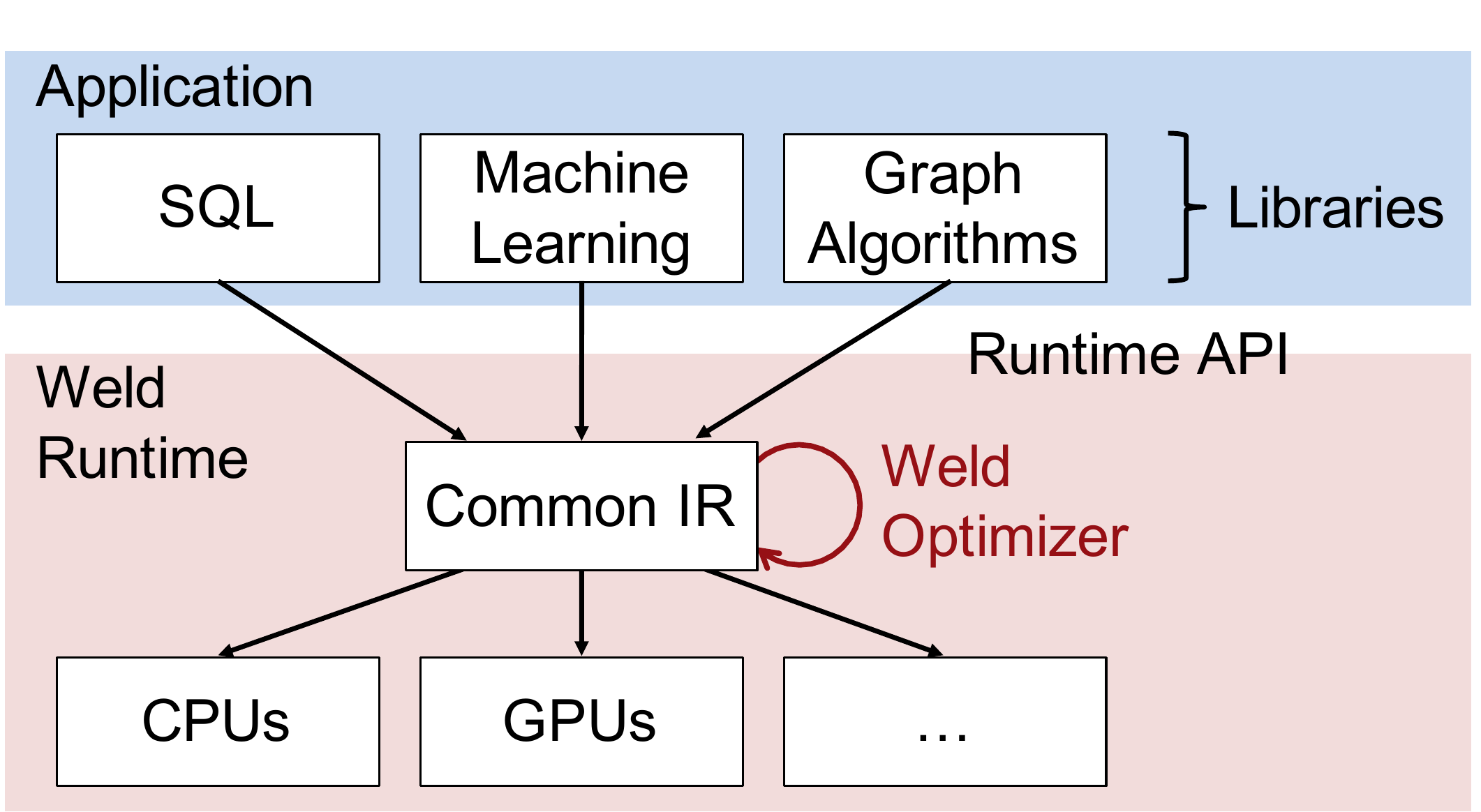}
  \setlength{\belowcaptionskip}{-10pt}
  \caption{\sn captures diverse data-parallel workloads using a
  common intermediate representation (IR) to emit efficient code for the
  end-to-end application on diverse hardware.}
  \label{fig:intro-overview}
\end{figure}

This paper proposes \sn, a novel interface and runtime that can optimize \emph{across}
data-intensive libraries and functions while preserving their user-facing APIs. \sn consists of
three key components (Figure~\ref{fig:intro-overview}).  First, \sn asks libraries to represent
their computations using a functional \emph{intermediate representation (IR)}.  This \il captures
the data-parallel structure of each function to enable rich data movement optimizations such as loop
fusion and tiling~\cite{loop-tiling}.  Libraries then submit their computations to \sn through a
lazily-evaluated \emph{runtime API} that can collect \il code from multiple functions before
executing it.  Finally, \sn's \emph{optimizer} combines these \il fragments into efficient machine
code for diverse parallel hardware. \sn's approach, which combines a unified \il with lazy
evaluation, enables complex optimizations \emph{across} independently written
libraries for the first time.

\sn's first component is its \il. We sought to design an \il that is both highly general (to
capture a wide range of data analytics computations) and amenable to complex optimizations (\eg loop
fusion, loop tiling, vectorization, and execution on diverse hardware). To this end, \sn uses
a small functional \il based on two concepts: nested parallel loops and an abstraction called
\emph{builders} for composing results in parallel. Builders are a hardware-independent abstraction
that specify \emph{what} result to compute (\eg a sum or a list) without giving a low-level
implementation (\eg atomic instructions), allowing for different implementations on different
hardware.
While \sn's \il imposes some limitations, such as lack of support for asynchronous computation, we
show that it is general enough to express relational, graph and machine learning workloads, and to
produce code with state-of-the-art performance for these tasks.

\sn's second component is a runtime API that uses lazy evaluation to capture work across function call and
library boundaries. Unlike interfaces such as OpenCL and CUDA~\cite{opencl,cuda}, which execute work
eagerly, \sn registers the work from multiple functions (even in different languages) and optimizes across them only when the program forces an evaluation (\eg before writing to disk).
The API supports accessing data in the application's memory without changing its format, allowing \sn to work against the native in-memory formats of common libraries such as Spark SQL and NumPy.

Finally, \sn's optimizer performs a wide range of optimizations on its \il, including loop fusion, loop tiling, and vectorization, to combine the \il fragments from different libraries into efficient machine code.
Although these optimizations are not novel, we show that they can be combined on \il fragments brought together by \sn's API to yield powerful optimizations across libraries that cannot be applied to the individual functions.
For example, in an application that filters data
using Spark SQL and applies a NumPy function to each row, \sn can vectorize the NumPy function
across rows, or even apply loop tiling~\cite{loop-tiling} across the two libraries.

We show that \sn can unlock order-of-magnitude speedups in data analytics applications, even when they use well optimized libraries.
We implemented a prototype of \sn with APIs in C, Java and Python, a full backend for multicore x86 CPUs, and a partial backend for GPUs.
We then integrated \sn into four common libraries: Spark SQL, NumPy, Pandas, and TensorFlow. 
In total, \sn can offer speedups of 3--29$\times$ in applications that use
multiple libraries, and 2.5--6.5$\times$ even in applications that use multiple
functions from the same library,
by minimizing data movement and generating efficient machine code.
Moreover, because \sn's \il is data-parallel, it can also parallelize the computations of single-threaded libraries such as NumPy and Pandas, yielding speedups of up to 180$\times$ when allowed to use more cores than the original computation with no additional programmer effort.
\sn's compiler is also competitive with code generators for narrower domains, such as HyPer~\cite{hyper} for SQL and XLA~\cite{xla} for linear algebra.
Finally, porting each library to use \sn
only required a few days of effort and could be done incrementally,
with noticeable benefits even when just a few common operators were ported.

To summarize, the contributions of this paper are:

\begin{denseenum}
\item A novel interface to enable cross-library and cross-function optimizations in
    data-intensive workloads using (1) a general intermediate
    representation based on loops and builders and (2) a lazily evaluated runtime API.
    
\item An optimizer that combines \sn \il fragments from disjoint libraries into efficient code for
    multicores and GPUs.
    
\item An evaluation of \sn using integrations with Pandas, NumPy, TensorFlow and Spark
    that shows \sn can offer up to 29$\times$ speedups in existing applications.
\end{denseenum}

\section{System Overview}
\label{sec:motivation}

\begin{figure}[t]
  \centering
  \includegraphics[width=\columnwidth]{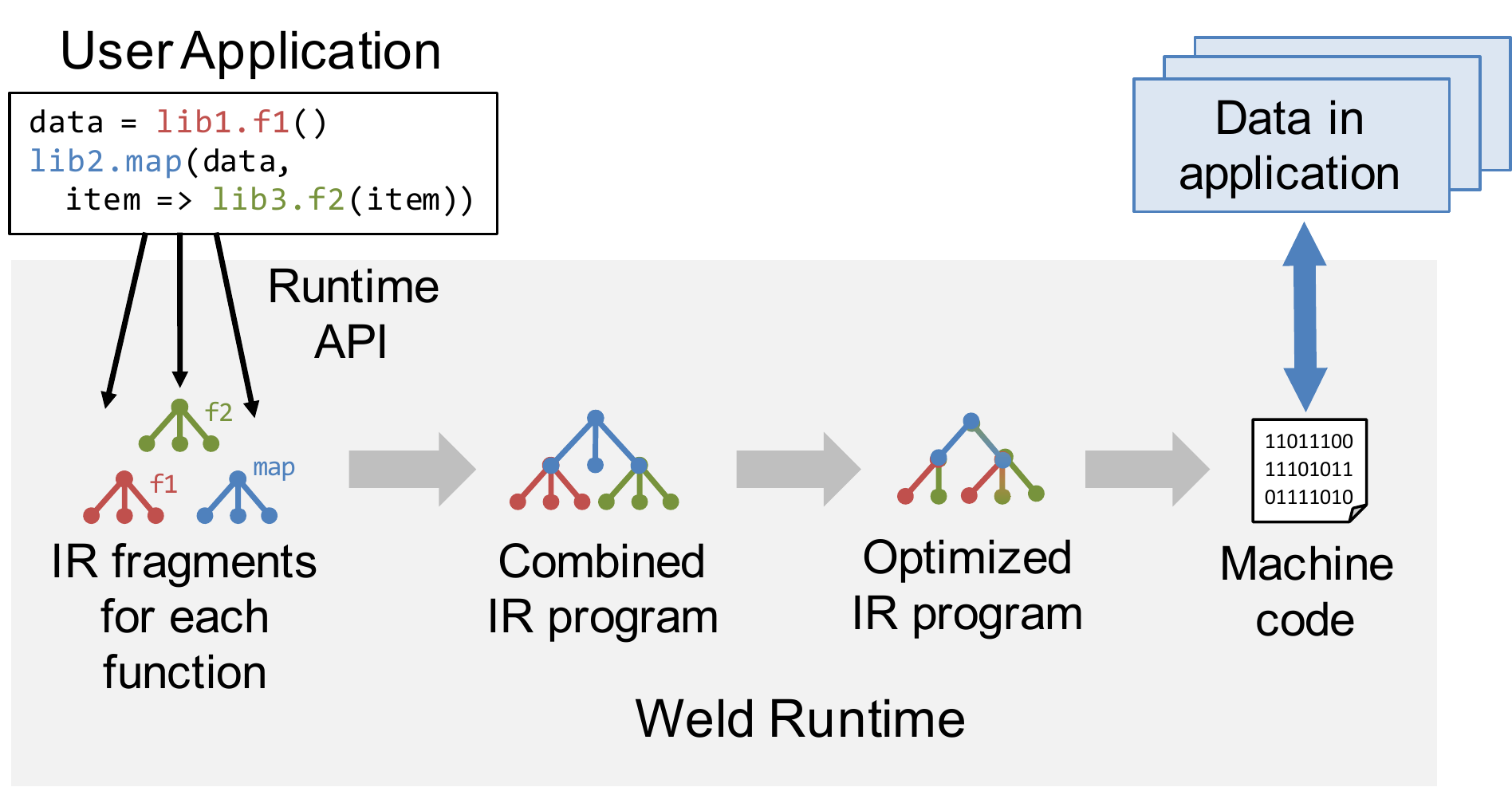}
  \setlength{\belowcaptionskip}{-10pt}
  \caption{Overview of \sn.  \sn collects fragments of \il code for each
    \sn-enabled library call and combines them into a single \il program.
    It then compiles this program to optimized code that runs on
    the application's in-memory data.
  }
  \label{fig:system-overview}
\end{figure}

Figure~\ref{fig:system-overview} shows an overview of \sn.
As described earlier, \sn has three components: a data-parallel \il for libraries to express work in, a lazy runtime API for submitting this work, and an optimizer that targets various hardware.
These components can be integrated into existing user-facing libraries.

In practice, we expect libraries to integrate \sn in two main ways.
First, many libraries, such as Pandas and NumPy, already implement their key functions in low-level languages such as OpenCL or C.
Developers can port individual functions to emit \sn code instead, and then automatically benefit from \sn's cross-function optimizations.
Libraries like NumPy and Pandas already have a compact in-memory representation for data (\eg NumPy arrays~\cite{numpy}), so \sn can work directly against their in-memory data at no extra cost.
\sn's design allows functions to be ported incrementally and offers notable speedups even if
only a few common operators are ported.

Second, some libraries, such as Spark SQL and TensorFlow, already implement code generation beneath a lazily evaluated API.
For these libraries, \sn offers both the ability to interface efficiently with other libraries,
and a simpler way to generate code.
For example, much of the complexity in code generators for databases involves operator
fusion logic: transforming a tree of operators into a single, imperative loop over the data~\cite{hyper,spark-wholestage}.
With \sn, each operator can emit a separate loop, independent of downstream operators; \sn
will then fuse these loops into a single, efficient program.

We note that \sn focuses primarily on \emph{data movement} optimizations for \emph{data-parallel}
operators from domains such as relational and linear algebra. These operators consume the bulk of
time in many applications by causing memory traffic, and benefit from co-optimization.
Domain-specific optimizations, such as reordering linear algebra expressions or ordering joins in
SQL, still need to be implemented within each library (and outside of \sn). 
In addition, \sn supports calling existing C functions for complex non-data-parallel code that
developers have already optimized.
Nonetheless, we show that \sn's data movement optimizations have a significant impact.

\subsection{A Motivating Example}
\label{sub:motivation-example}

We illustrate the benefit of \sn in a data science workflow adapted from a
tutorial for Pandas and NumPy~\cite{pandas-workload}. Pandas and NumPy
are two popular Python data science libraries: Pandas provides a ``dataframe'' API for manipulating
data in a tabular format, while NumPy provides fast linear algebra operators.  Both Pandas and NumPy offer optimized operators, such as data filtering and vector addition, written in C or Cython.
However, workloads that combine these operators still experience substantial
overhead from materializing intermediate results.

\begin{figure}[t]
  \centering
  \setlength{\belowcaptionskip}{-5pt}
  \includegraphics[width=\columnwidth]{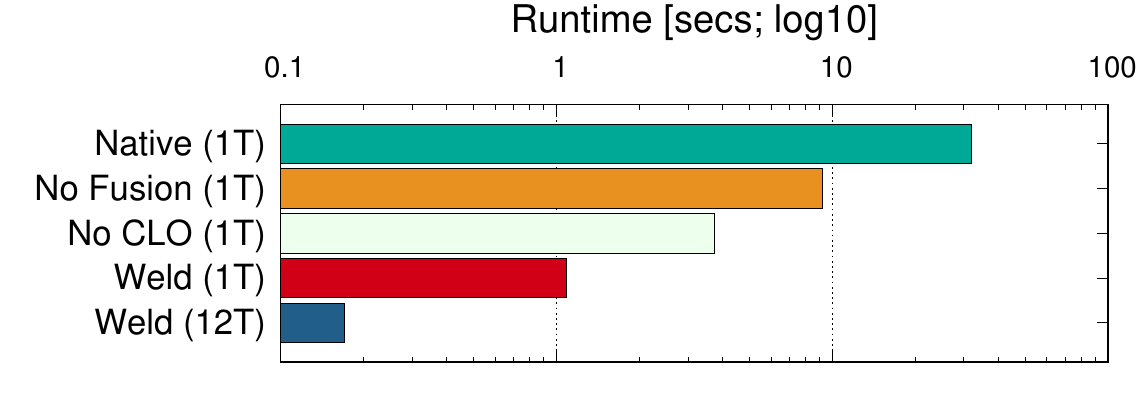}
  \caption{Performance of a data science workflow (log scale) in unmodified Pandas and NumPy
  (where only individual operators are written in C), \sn without loop fusion,
  \sn without cross-library optimization (CLO),
  \sn with all optimizations enabled, and \sn with 12 threads.
  }
  \label{fig:overview-example}
  \squeezeup
\end{figure}

Our workload consists of filtering large cities out of a
population-by-cities dataset, evaluating a linear model using features in the dataframe to
compute a crime index, and then aggregating these crime indices into a total crime index.
It combines relational operators from Pandas with vector arithmetic
operators from NumPy.
Figure~\ref{fig:overview-example} shows its performance on a 6 GB dataset.
Porting each operator to run on \sn yields a 3$\times$ speedup (shown in the No Fusion bar) due to
\sn's more efficient, vectorizing code generator.
Enabling \sn's loop fusion optimization then leads to a further 2.8$\times$ speedup \emph{within}
each library, and an additional 3.5$\times$ speedup when enabled \emph{across} libraries.
This gives \sn a total 29$\times$ speedup on a single thread, largely due to this data movement
optimization (\sn 1T bar).
Finally, Pandas and NumPy are single-threaded, but \sn can automatically parallelize its generated
code without any change to the user application. Enabling multithreading gives a further 6.3$\times$
speedup on 12 cores, at which point \sn saturates the machine's memory bandwidth, for a total
of 187$\times$ speedup versus single-core NumPy and Pandas.

\subsection{Limitations}
\label{subsec:limitations}

\sn's interface and implementation have several limitations.
First, \sn currently only aims to accelerate single-machine code in a shared-memory environment (\eg
multicore CPU or GPU). Its \il includes shared-memory operations such as
random lookups into an array, which are difficult to implement efficiently in a distributed setting.
Nonetheless, \sn
can also be integrated into distributed systems such as Spark to accelerate each node's local
computations.  These distributed frameworks are often CPU- or memory-bound~\cite{ousterhout2015,sparksql,tupleware}.
We show in \S~\ref{sec:eval} that by optimizing computation on each node, \sn can
accelerate Spark applications by 5-10$\times$.

Second, \sn's \il cannot express asynchronous computations where threads race to update a
result~\cite{hogwild}; all \sn programs are race-free.
It also lacks constructs to let programmers control locality (\eg pinning data to a CPU socket).

Third, \sn executes computations lazily, which makes programs harder to debug.
Fortunately, lazy APIs are becoming very common in performance-sensitive systems such as LINQ~\cite{linq}, Spark and TensorFlow, and we believe that programmers can use similar debugging techniques
(\eg printing an intermediate result).

Finally, \sn requires integration into libraries in order to speed up applications.
As discussed at the beginning of \S\ref{sec:motivation}, we believe that this is worthwhile for several reasons.
First, many libraries already write their performance-sensitive operators in C or OpenCL; \sn offers a higher-level (functional) 
interface to write code that is more hardware-independent.
Second, \sn enables significant speedups \emph{even within a single library}, creating incentives for individual libraries to use it.
Finally, \sn can be integrated incrementally and still offer substantial speedups, as we show in \S\ref{sub:eval-incremental}.


\section{\sn's Intermediate Representation}
\label{sec:language}

\setlength\extrarowheight{2pt}
\lstset{language=NVL}

Libraries communicate the computations they perform to \sn using a data-parallel intermediate representation
(\il).  This component of the \sn interface determines both which workloads can run on \sn and which
optimizations can easily be performed.
To support a wide range of data-intensive workloads, we designed \sn's \il to meet three goals:
\begin{denseenum}
  \item Generality: we wanted an \il that could express diverse data analytics
  tasks (\eg relational and linear algebra), as well as \emph{composition} of these
  tasks into larger programs. 
  \item Ability to support optimizations, especially for data movement optimizations such as
  loop fusion and loop tiling.
  \item Parallelism: we wanted the \il to be explicitly parallel so that \sn can automatically
  generate code for modern parallel hardware, \eg multicores and GPUs.
\end{denseenum}

To meet these goals, we designed a small \il inspired by monad
comprehensions~\cite{monad-comprehensions}, similar to functional languages but
operating at a lower level that makes it easier to express fusion optimizations.

\subsection{Data Model}

\sn's basic data types are scalars (\eg \lstinline$int$ and
\lstinline$float$), 
variable-length vectors (denoted \lstinline$vec[T]$ for a type \lstinline$T$),
structures (denoted \lstinline${T1,T2,...}$), and dictionaries
(\lstinline$dict[K,V]$).
These types are nestable to support more complex data.
We chose these types because they appear commonly in data-intensive applications and
in low-level data processing code (\eg dictionaries are useful to implement database joins).

\subsection{Computations}
\label{sub:operators}

\sn's \il is a functional, expression-oriented language. It contains basic operators for
arithmetic, assigning names to values, sequential \lstinline$while$ loops, and collection lookups. It also supports calling external functions in C.

In addition, the \il has two parallel constructs: a parallel \lstinline$for$ loop and a
hardware-independent abstraction for constructing results in parallel called a \emph{builder}.
Parallel loops can be nested arbitrarily, which allows complex function composition. Each loop can
merge values into multiple builders; for example, a single loop can merge values into one
builder to produce a sum and another to produce a list.

\sn includes multiple types of builders, shown in Table~\ref{table:buildertypes}.  For example, a
\lstinline$vecbuilder[T]$ takes values of type \lstinline$T$ and builds a vector of merged
values. A \lstinline$merger[T,func,id]$ takes a commutative function and an identity value and
combines values of type \lstinline$T$ into a single result.

Builders support three basic operations. \lstinline$merge(b, v)$ adds a new value \lstinline$v$ into
the builder \lstinline$b$ and returns a new builder to represent the result.\footnote{In practice, some mutable state
will be updated with the merged value, but \sn's functional \il treats all values as immutable, so we
represent the result in the \il as a new builder object.}
Merges into
builders are associative, enabling them to run in parallel.  \lstinline$result(builder)$
destroys the builder and returns its final result: no further operations are allowed on it after
this.
Finally, \sn's parallel \lstinline$for$ loop is also an operator that consumes and returns builders.
\lstinline$for(vectors, builders, func)$ applies a function of type \lstinline$(builders, index, elem) => builders$ 
to each element of one or more vectors in parallel, then returns the updated builders.
Each call to \lstinline$func$ receives the index of the corresponding element and a struct with the
values from each vector.
The loop can also optionally take a start index, end index and stride for each input vector to
express more complex strided access patterns over multidimensional arrays (\eg matrices).

\begin{table}[t!]
\centering
\footnotesize
\begin{tabular}{|p{1.15in}|p{1.75in}|}
\hline
\multicolumn{2}{|c|}{\textbf{Builder Types}}                                                                                                                                      \\ \hline
{\lstinline$vecbuilder[T]$}           & {Builds a \lstinline$vec[T]$ by appending merged values of type \lstinline$T$}                                      \\ \hline
{\lstinline$merger[T,func,id]$}     & {Builds a value of type \lstinline$T$ by merging values using a commutative function \lstinline$func$ and an identity value \lstinline$id$} \\ \hline
{\lstinline$dictmerger[K,V,func]$}        & {Builds a \lstinline$dict[K,V]$ by merging \lstinline${K,V}$ pairs using a commutative function}                              \\ \hline
{\lstinline$vecmerger[T,func]$} & {Builds a \lstinline$vec[T]$ by merging \lstinline${index,T}$ elements into specific cells in the vector using a commutative function}                 \\ \hline
{\lstinline$groupbuilder[K,V]$}       & {Builds a \lstinline$dict[K,vec[V]]$ from values of type \lstinline${K,V}$  by grouping them by key}                        \\ \hline
\end{tabular}
\setlength{\belowcaptionskip}{-10pt}
\caption{Builder types available in \sn.}
\label{table:buildertypes}
\end{table}

\begingroup
\begin{lstlisting}[language=NVL]
// Merge two values into a builder
b1 := vecbuilder[int];
b2 := merge(b1, 5);
b3 := merge(b2, 6);
result(b3)   // returns [5, 6]

// Use a for loop to merge multiple values
b1 := vecbuilder[int];
b2 := for([1,2,3], b1, (b,i,x) => merge(b, x+1));
result(b2)   // returns [2, 3, 4]

// Loop over two vectors and merge results only
// on some iterations.
v0 := [1, 2, 3];
v1 := [4, 5, 6];
result(
  for({v0, v1},
     vecbuilder[int],
     (b,i,x) => if(x.0 > 1) merge(b, x.0+x.1) else b
)) // returns [7, 9]
\end{lstlisting}
\captionof{lstlisting}{Examples of using builders.}
\label{listing:builders}
\endgroup

\sn's \lstinline$for$ loops can be nested arbitrarily, enabling \sn to express irregular parallelism (\eg graph algorithms) where different iterations of
the inner loop do different amounts of work.

Finally, \sn places two restrictions on the use of builders for efficiency. First, each builder must be
consumed (passed to an operator) exactly once per control path to prevent having multiple values
derive from the same builder, which would require copying its state. Formally, builders are a
linear type~\cite{walker-types}.  Second, functions passed to \lstinline$for$ must return builders
derived from their arguments. These restrictions let \sn's compiler safely implement builders using
mutable state.


\subsection{Higher-Level Operators}

To aid developers, \sn also contains macros that implement
common functional operators such as \lstinline$map$, \lstinline$filter$ and \lstinline$reduce$
using builders, so that library developers familiar with functional APIs can concisely express their
computations. These operators are all map into loops and builders. For
example, the second snippet in Listing~\ref{listing:builders} implements a \lstinline$map$ over
a vector.

\subsection{Why Loops and Builders?}

Given the broad use of functional APIs such as MapReduce and Spark, a strawman design for an \il
might have used these functional operators as the core constructs rather than loops and builders.
Unfortunately, this design prevents expressing many common optimizations in the \il.
For example, consider Listing~\ref{listing:nofuse}, which runs two operations on the same input vector:

\begingroup
\begin{lstlisting}[language=NVL]
data := [1,2,3];
r1 := map(data, x => x+1);
r2 := reduce(data, 0, (x, y) => x+y)
\end{lstlisting}
\captionof{lstlisting}{A \lstinline$map$ and
\lstinline$reduce$ over the same input vector.}
\label{listing:nofuse}
\endgroup

Even though these \lstinline$map$ and \lstinline$reduce$ operations can be computed in a
shared pass over the data, no operator akin to {\lstinline$mapAndReduce$} exists that computes
both values in one pass. Richer optimizations such as loop tiling are even
harder to express in a high-level functional \il.
By exposing a single loop construct that can update
multiple builders, patterns like the above can easily be fused into programs such as
Listing~\ref{listing:horizontalfusion}.

\begingroup
\begin{lstlisting}[language=NVL]
data := [1,2,3];
result(
  for(data, { vecbuilder[int], merger[+,0] },
     (bs, i, x) => { merge(bs.0, x+1), merge(bs.1, x) }
)) // returns {[2,3,4], 6}
\end{lstlisting}
\captionof{lstlisting}{\lstinline$for$ operating over multiple builders to produce both a vector and an aggregate in one pass.}
\label{listing:horizontalfusion}
\endgroup

\subsection{Generality of the \il}
\label{sub:generality}

\sn's parallel loop and builders can express all of the functional operators in systems such as
MapReduce, LINQ and Spark, as well as relational algebra, linear algebra and other data-parallel operations.
Given the wide range of algorithms implemented over these APIs~\cite{cacm-spark,dryadlinq}, we believe
\sn can benefit many important workloads.
Our evaluation shows workloads from each of these domains. 
As discussed in \S~\ref{subsec:limitations}, the current \il does not capture asynchrony.

\section{Runtime API}
\label{sec:frontends}

\lstset{language=Python}

The second component of \sn is its runtime API.  Unlike interfaces like OpenCL and CUDA, \sn uses a
\emph{lazy} API to construct a computation graph. Library functions use the API to compose fragments
of \il code (perhaps across libraries) and provide access to data in the application.
The API uses the \il fragments and data to build a DAG of computations.
When libraries evaluate a computation, \sn fuses the graph into a single \il program,
optimizes it and and executes it.
We show examples of the API in Python here, it also supports Java and C.

Consider the program in Listing~\ref{listing:pythonex} as a motivating example, which uses
the \lstinline$itertools$ library's \lstinline$map$ function to iterate over a set of vectors and
apply \lstinline$numpy.dot$ to each one:

\begingroup \begin{lstlisting}[language=Python]
scores = itertools.map(vecs, lambda v: numpy.dot(v, x))
print scores
\end{lstlisting}
\captionof{lstlisting}{A Python program that combines library calls.}
\label{listing:pythonex}
\endgroup

A standard call to \lstinline$itertools.map$ would treat \lstinline$numpy.dot$ as a black box and
call it on each row.  If both libraries use \sn, however, the result \lstinline$scores$ is instead
an object encapsulating a \sn program capturing \emph{both} functions. \sn evaluates this object
only when the user calls \lstinline$print$. Before evaluation, \sn will optimize the \il code for
the entire workflow, enabling optimizations that would not make sense in either function on its own. For
example, \sn can tile the loop to reuse blocks of the \lstinline$x$ vector across multiple rows of
\lstinline$v$ for cache efficiency.

\newcommand{\snobj}{\lstinline$WeldObject$\xspace}
\newcommand{\snobjs}{\lstinline$WeldObject$s\xspace}
\newcommand{\snobjnew}{\lstinline$NewWeldObject$\xspace}
\newcommand{\snobjfree}{\lstinline$FreeWeldObject$\xspace}

\newcommand{\snvalue}{\lstinline$WeldResult$\xspace}
\newcommand{\snvalues}{\lstinline$WeldResult$s\xspace}
\newcommand{\snvaluefree}{\lstinline$FreeWeldResult$\xspace}

\begin{table}[t!]
\centering
\small
\begin{tabular}{|p{1.15in}|p{1.75in}|}
\hline
\multicolumn{2}{|c|}{\textbf{API Summary}}                                                                                                                          \\ \hline
\lstinline$NewWeldObject(data, type, encoder)$      & Creates a \snobj wrapping the given
    in-memory \lstinline$data$ and giving it \sn type \lstinline$type$.
    The \lstinline$encoder$ object implements marshaling
    (\S\ref{subsec:api-encoding}). \\ \hline
\lstinline$NewWeldObject(deps, expr, encoder)$      & Creates a \snobj with the given
    dependencies (other \snobjs), a \sn \il expression, and an \lstinline$encoder$ for the resulting type.
   \\ \hline
\lstinline$GetObjectType(o)$                         & Returns the \sn type of the \snobj \lstinline$o$ (\eg \lstinline$vec[int]$).                                              \\ \hline
\lstinline$Evaluate(o)$                      & Evaluates the \sn program and returns a
    result.                                       \\ \hline
\lstinline$FreeWeldObject(o)$                        & Frees memory for this \snobj.                                                                   \\ \hline
\lstinline$FreeWeldResult(v)$                         & Free a \snvalue.                                                                                \\ \hline
\end{tabular}%
\setlength{\belowcaptionskip}{-10pt}
\caption{A summary of the \sn API.}
\label{table:apidescription}
\end{table}

\subsection{API Overview}
\label{sub:api}

Developers integrate \sn into their libraries using an interface called \snobj, which represents
either a lazily evaluated sub-computation or external data in the application. A \snobj
may depend on other \snobjs (possibly from other libraries), forming a DAG for the whole
program where leaves are external data. Table~\ref{table:apidescription} summarizes \sn's API.

Developers create \snobjs using the \snobjnew call. This call has two variants: one to encapsulate
external data dependencies in the application and one to encapsulate sub-computations and dependencies
\emph{among} \snobjs. To encapsulate an external data dependency, developers pass as arguments a pointer to
the data dependency, the \sn type of the dependency (\eg \lstinline$vec[int]$), and an
\emph{encoder} for marshaling between native library formats and \sn-compatible objects.
\S\ref{subsec:api-encoding} discusses encoders in detail.

To encapsulate sub-computations and dependencies with other \snobjs, developers pass a list of
dependent \snobjs (\lstinline$deps$), a \sn \il expression representing the computation, and an
encoder for the expected type of the \snobj. If the library evaluates the \snobj, this encoder is
used to marshal data returned by \sn to a native-library format.  The provided \il expression must
depend only on the dependencies declared in \lstinline$deps$.
Listing~\ref{listing:api-code-example} shows an example
function for computing the square of a number using \sn's API.

\begingroup
\begin{lstlisting}[language=Python]
def square(self, arg):
  # Programatically construct an IR expression.
  expr = weld.Multiply(arg, arg)
  return NewWeldObject([arg], expr)
\end{lstlisting}
    \captionof{lstlisting}{A simple function that squares an argument, \lstinline$arg$, passed in as
    a \snobj.
    }
    \label{listing:api-code-example}
\endgroup

The \lstinline$Evaluate$ API call evaluates a \snobj instance and returns a result.  Libraries can
choose when to evaluate an object in several ways.  In our integrations with Python libraries, we
used methods that save or print the object (\eg the \lstinline$__str__$ method to convert it to a
string) as evaluation points to introduce lazy evaluation behind the library's existing
API. Systems like Spark and TensorFlow already have lazy APIs with well-defined evaluation points.
\lstinline$Evaluate$ returns a special handle
called \snvalue that can be checked for failure or queried for a pointer to the returned data.

Library developers manage the lifecycle of a \snobj manually after instantiation. The \snobjfree
call deletes an instance of \snobj by freeing its internal state; this call \emph{does not} free
data dependencies or child \snobj instances in other libraries.  In languages with automatic memory
management, such as Python, developers can add the \snobjfree call in their class's destructor.

\subsection{Marshalling Data}
\label{subsec:api-encoding}

\sn specifies a standard binary format for its basic data types that allows it to operate over
existing in-memory data. Specifically, scalar types (\lstinline$int$, \lstinline$float$, \etc) and
structs follow C packed structure layout, and vectors \lstinline$vec[T]$ are represented as an
\lstinline${int64,T*}$ structure. Dictionaries and builders cannot be passed into \sn from outside
in our current implementation.

Library developers provide \emph{encoders} to map types in their native programming languages (\eg
Python) to \sn-usable data and vice-versa. The encoder interface is a single function
\lstinline$encode$. When used with \snobjnew to declare a data dependency, this function maps an
object in the library's native language and returns data understood by \sn's runtime. For example,
an encoder for NumPy arrays would take in a NumPy object and extract a pointer to its internal data
array, which already happens to be a packed array of primitive types in NumPy~\cite{numpy}. When
used with \snobjnew to declare a sub-computation, the function takes a pointer to data in \sn's
format and returns an object in a library format.

\subsection{Memory Management}
\label{subsec:api-memory}

Values in \sn are considered either \emph{owned by library} or \emph{owned by \sn}.  Input data
is always owned by libraries. These data are neither mutated nor freed by \sn.  Values
that \sn allocates during execution are owned by \sn, as are results of \lstinline$Evaluate$
(\snvalues). Decoders can either point to these values directly when wrapping them into library-specific
objects (as long as the library does not free the corresponding \snvalue), or they can copy data.

\sn can also allocate memory during the execution of an \il program (\ie temporary values that are
not part of the final result).  The runtime frees this memory after each call to
\lstinline$Evaluate$, preserving only the objects needed for the resulting \snvalue.  The
\lstinline$Evaluate$ function takes a memory limit as an argument to prevent unbounded
allocation, which is useful when integrating \sn into a larger engine such as Spark SQL.

\begin{figure}[t]
  \centering
  \includegraphics[width=0.9\columnwidth]{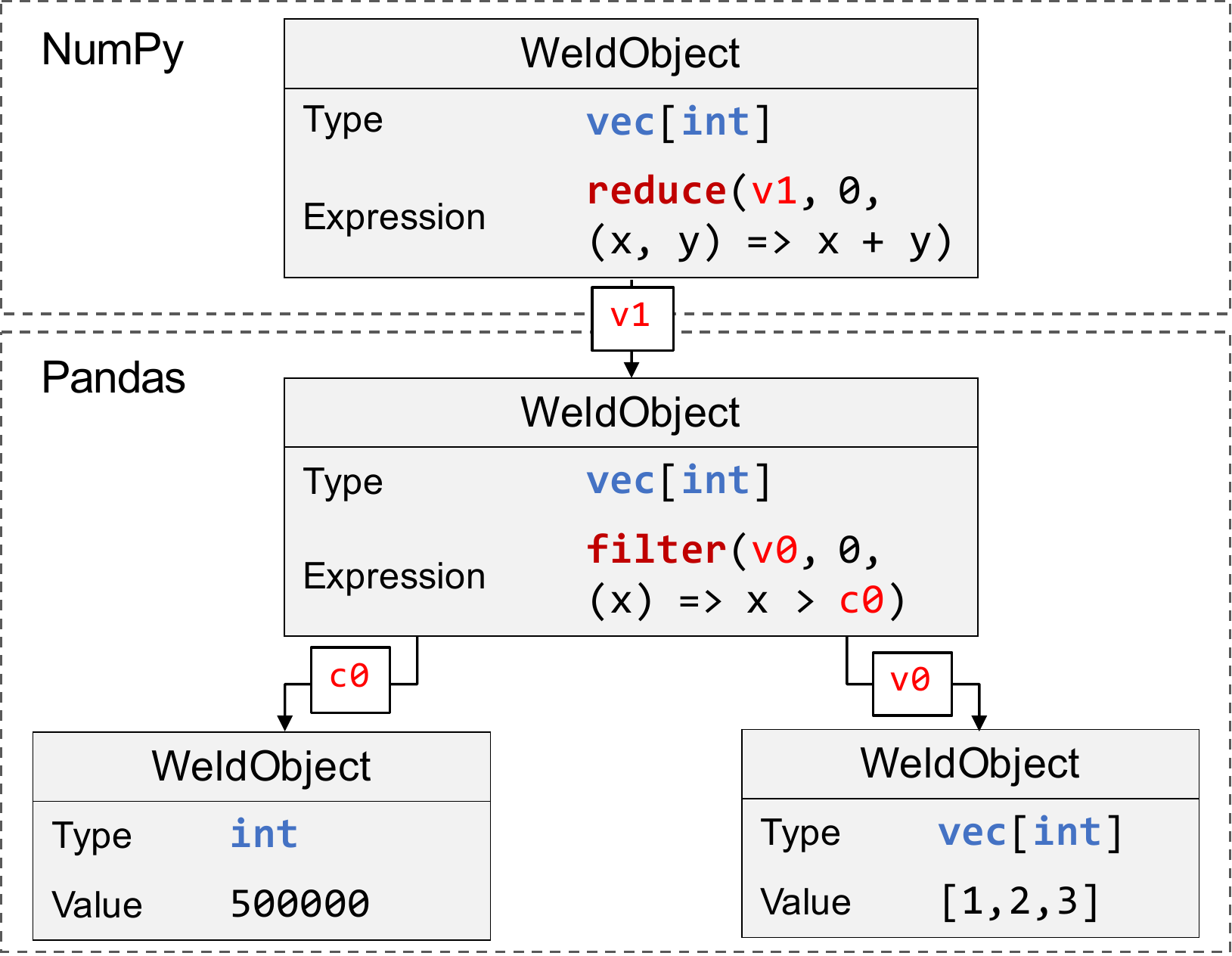}
  \setlength{\belowcaptionskip}{-10pt}
    \caption{The example from Listing~\ref{listing:runtime-example-python} as a computation graph.}
    \label{fig:api-example}
\end{figure}

\subsection{User Defined Functions (UDFs)}
\label{sub:pyudfs}

Libraries such as Spark and \lstinline$itertools$ can take functions as arguments.
To implement these, a \snobj's \il expression can also represent a function, with dependencies pointing to variables in its closure.
In order to make it easy to pass functions into \sn, we implemented a basic UDF translator for Python that
walks Python abstract syntax trees (ASTs) to convert them to \sn \il, based on techniques in existing systems~\cite{dandelion,tupleware,pydron}.
Listing~\ref{listing:udf-example-python} shows
an example; the \lstinline$@weld$ annotation provides a type signature for the function in \sn.

\begingroup
\begin{lstlisting}[language=Python]
# Produces (x: int) => x + 1 in Weld IR
@weld("(int) => int")
def increment(x): return x + 1
\end{lstlisting}
\captionof{lstlisting}{Python UDF with a \sn type annotation.}
\label{listing:udf-example-python}
\endgroup

\subsection{Example: Combining NumPy and Pandas}
\label{subsec:runtime-example}

\sn's API enables optimizations even across independent libraries. We illustrate this with
the function in Listing~\ref{listing:runtime-example-python}, which
uses the Python Pandas library and NumPy to compute the total population of all cities with over
500,000 residents. NumPy represents data in C arrays wrapped in Python objects. Pandas uses data
frames~\cite{pandas}, which are tables with named columns that are also encoded as NumPy arrays.

\begingroup
\begin{lstlisting}[language=Python]
def large_cities_population(data):
  filtered = data[data["population"] > 500000]
  sum = numpy.sum(filtered)
  print sum
\end{lstlisting}
\captionof{lstlisting}{A sample Python program using Pandas and NumPy.}
\label{listing:runtime-example-python}
\endgroup

In the native Pandas and NumPy libraries, this code causes two data scans:
one to filter out values greater than 500,000 and one to sum the values.
Using \sn, these scans can be fused into a single
loop and the sum can be computed ``on the fly.'' The loop
also benefits from optimizations such as vectorization.

To enable using \sn for this program, we must extend the \lstinline$DataFrame$ object in Pandas to
return lazily evaluated \snobjs. We must also provide a \il fragment for the
\lstinline$>$ operator on \lstinline$DataFrame$ columns and the \lstinline$numpy.sum$ function.

Listing~\ref{listing:runtime-example-impl} shows the \sn expressions for implementations
of each of these functions. Each implementation takes either a data
dependency (\eg a NumPy array) or \snobj as input, and incrementally builds a
\sn computation graph by returning another composable \snobj instance.

\begingroup
\begin{lstlisting}[language=Python]
# DataFrame > filter
# Input: Vector v0, constant c0
filter(v0, (x) => x > c0)

# numpy.sum
# Input: Vector v0
reduce(v0, 0, (x, y) => x + y)
\end{lstlisting}
\captionof{lstlisting}{Pandas and NumPy functions using \sn.}
\label{listing:runtime-example-impl}
\endgroup

After porting these operators to \sn, the libraries can use the API to lazily compose a
computation graph for the full workload without any modifications to the user's code.
Listing~\ref{listing:runtime-example-fused} shows the final fused \sn expression of the variable
\lstinline$sum$ (now a \snobj) before it is printed.  Calling \lstinline$print$ on this instance
invokes the \lstinline$evaluate$ method to compute a result. Figure~\ref{fig:api-example} shows the
computation graph for the workload.

\begingroup
\begin{lstlisting}[language=NVL]
reduce(filter(v0, (x) => x>500000), 0, (x,y) => x+y)
\end{lstlisting}
\captionof{lstlisting}{The combined \sn program.}
\label{listing:runtime-example-fused}
\endgroup

After optimization, this function becomes a single parallel loop using
builders, as shown in Listing~\ref{listing:runtime-example-final}:

\begingroup
\begin{lstlisting}[language=NVL]
result(
  for(v0, merger[+,0],
    (b, x) => if (x > 500000) merge(b, x) else b))
\end{lstlisting}
\captionof{lstlisting}{The \sn program after loop fusion.}
\label{listing:runtime-example-final}
\endgroup

The optimizer then applies optimizations such as predication and
vectorization here; for clarity, we omit these above.


\section{Optimizer and Hardware Backends}
\label{sec:implementation}

\sn's optimizer combines \il fragments composed by different libraries through
the runtime API into efficient machine code for parallel hardware.
The optimizer starts by applying some general optimization rules in the \il
(\ie outputting new code in the same \il).
It then passes the code to hardware-specific backends.
Although \sn uses standard compiler optimizations,
our contribution is to show that its compilation strategy produces efficient
code for libraries combined using \sn's runtime API, even when these libraries
implement their functions using \sn's \il in isolation.

\paragraph{\il-level Optimizations.}

We modeled our optimizer after LLVM~\cite{llvm}, which applies a wide range of
hardware-independent optimizations at the \il level and then passes a
largely optimized program to a hardware backend.
This approach is powerful because it allows composing passes at the \il level in
arbitrary ways, and it fits well with \sn's loop-and-builder based \il
(\S\ref{sec:language}), which can express fused and tiled code within the same \il
as the original program.

\begin{table}[t!]
\centering
\small
\begin{tabular}{|p{0.73in}|p{2.17in}|}
\hline
\multicolumn{2}{|c|}{\textbf{Optimizations Passes}} \\ \hline
Loop Fusion & Fuses adjacent loops to avoid materializing
	intermediate results when the output of one loop is used as the input of another. Also fuses
	multiple passes over the same vector. \\ \hline
Size Analysis & Infers the size of output vectors statically. \\ \hline
Loop Tiling & Breaks nested loops into blocks to exploit caches by reusing values faster~\cite{loop-tiling}. \\ \hline
Vectorization \& Predication & Transforms loops with simple inner bodies to use vector instructions.
Branches inside the loop body are transformed into unconditional select instructions (predication).
 \\ \hline
Common Subexpression Elimination & Transforms the program to not run the same computation multiple times. \\ \hline
\end{tabular}%
\setlength{\belowcaptionskip}{-10pt}
\caption{Optimization passes implemented in \sn.}
\label{table:optimization}
\end{table}

We implemented several different optimization passes at the \il level, shown in Table~\ref{table:optimization}.
These passes are implemented using pattern-matching rules on sub-trees of the
abstract syntax tree (AST).
\sn applies passes in a static order, with rules at each level applied repeatedly
until the AST no longer changes. Specifically, we apply
the loop fusion transformations first, then size analysis, then loop tiling, then
vectorization and finally common subexpression elimination.

\sn's functional \il makes standard optimizations significantly easier to apply than to C or LLVM.
For example, common optimizations such as vectorization are hard to apply in C or LLVM because of
pointer aliasing (determining whether two pointers could refer to overlapping addresses).  In contrast,
\sn's immutable data values and ``write-then-read'' builders (which have a separate ``write-only''
phase followed by a read-only phase after computing the result) make it straightforward to transform
sub-trees of the AST in isolation.  As we show in \S\ref{sec:eval}, \sn's optimization
passes produce efficient code across a variety of domains, even when starting with disparate \il
fragments combined at runtime using \sn's API.

\paragraph{Multicore x86 Backend.} \sn's CPU backend emits explicitly multithreaded and
vectorized code for each program. We use LLVM~\cite{llvm} to generate code, and vectorize the code explicitly in our compiler to target AVX2~\cite{avx2} instructions. The backend could in principle run on non-x86 architectures supported by LLVM, but we have only evaluated it on x86.

At runtime, the generated code links with a multicore work-stealing execution engine inspired by Cilk~\cite{cilk}, which schedules work for each parallel \lstinline$for$ loop dynamically. 
This allows \sn to support irregular parallelism.
\eat{
The engine uses a modified work-stealing strategy tailored to \sn.
It represents a \sn program as a task graph with each outer \lstinline$for$
loop represented as a single task that depends on all loops and statements that come before it.
A loop can only execute after its dependencies complete.
The engine creates a worker thread on each core
and a task queue for each worker thread, where a ``task" is a range of loop iterations. 
Initially, all iterations of the first loop are given to the first worker. A worker steals tasks from other workers' queues when idle to load balance work. An executing
worker creates a new task with half its remaining iterations for the current loop when it observes
that its own task queue is empty.
}
In code with nested loops, new tasks are created by splitting up the outermost loop that still has more than one iteration remaining.
This policy ensures that expensive outer loops will be split across cores, but smaller inner loops can often stay on the same core without incurring task creation overhead.

We implemented the data structures within the backend, including builders, using standard multithreaded programming techniques such as cache line padding to reduce false sharing.
Multicore builders, including the \lstinline$merger$, are largely implemented with per-core copies that are aggregated when \lstinline$result$ is called. Other builder implementations are possible
and potentially faster in certain cases, as described in \S\ref{sub:eval-builders}.

\paragraph{Partial GPU Backend.}
Our GPU backend is built on top of OpenCL and supports the \lstinline$merger$, \lstinline$vecmerger$
and \lstinline$vecbuilder$ builders. \sn's
optimization passes fuse computations into a single OpenCL kernel that we then submit to the GPU.

In order to obtain reasonable performance on the GPU, we had to implement one more \il transformation rule specific to the GPU backend. Since GPUs cannot easily support dynamic parallelism at runtime, we use un-nesting~\cite{blelloch1988compiling} to transform a nested parallel program into a regular one.
In addition, our implementations of builders on the GPU are different.
For example, we use a parallel aggregation tree to combine intermediate results for \lstinline$merger$ and \lstinline$vecmerger$ efficiently across thousands of GPU cores.

Currently, programs in \sn must either execute \emph{completely} on the CPU or on the GPU. We do not yet support partial offloading of computation to a coprocessor.

\section{Library Integrations}
\label{sec:integrations}

To evaluate \sn, we integrated it into four popular libraries:
Spark SQL, TensorFlow, NumPy, and Pandas. 
Each integration required some up front ``glue code'' for marshalling data and enabling lazy
evaluation (if the library was eagerly evaluated), as well as code for each ported operator.
Overall, we found the integration effort to be modest across the board, as shown in
Table~\ref{tab:integration-effort}. Each library required 100--900 lines of glue code and an
additional 5--85 lines of code per operator; operators can be added incrementally and.
interoperate with native operators in each library.

\begin{table}[t]
\centering
\small
\begin{tabular}{|c|c|c|c|}
\hline
\textbf{Library}  & \textbf{Glue Code LoC} & \textbf{Per-Operator LoC}                                       \\ \hline
NumPy                         & Py: 84, C++: 24              & avg: 16,  max: 50 \\ \hline
Pandas                        & Py: 416, C++: 153            & avg: 22,  max: 64 \\ \hline
Spark SQL                   & Py: 5, Scala: 300            & avg: 23, max: 63 \\ \hline
TensorFlow                    & Py: 175, C++: 652            & avg: 22, max: 85 \\ \hline
\end{tabular}
  \setlength{\belowcaptionskip}{-10pt}
  \caption{
  Number of lines of code in our library integrations.
  }
  \label{tab:integration-effort}
\end{table}

\paragraph{Spark SQL.}
\label{subsec:sparksql-intergration}

\sn's integration with Spark SQL~\cite{sparksql} accelerates its local computations on each node. If
performance is bounded by local resources~\cite{ousterhout2015}, \sn accelerates queries even in
a distributed setting.  Spark SQL already has a lazy API to build an operator graph, and already
performs Java code generation using a similar technique to HyPer~\cite{hyper}, so porting this
framework was straightforward: we only needed to replace the emitted Java bytecode with \sn \il via
\sn's API.  Spark SQL's existing Java code generator uses complex logic~\cite{spark-wholestage} to
directly generate imperative loops for multiple chained operators because the Java compiler cannot
perform these optimizations automatically. In contrast, our \sn port emits a separate \il fragment
for each operator without considering its context, and \sn automatically fuses these loops.

\paragraph{TensorFlow.}
\label{subsec:tensorflow-intergration}

Like Spark SQL, TensorFlow~\cite{tensorflow} also has a lazily evaluated API that generates a data
flow graph composed of modular operators.  Our integration with TensorFlow required two components:
\emph{(i)} a user-defined \code{WeldOp} operator that runs an arbitrary \sn expression, and
\emph{(ii)} a graph transformer that replaces a subgraph of the TensorFlow data flow graph with an
equivalent \code{WeldOp} node. Before execution, the transformer searches the original data flow
graph for subgraphs containing only operators that are understood by our port, and replaces each
such subgraph with a \code{WeldOp} node for their combined expression, relying on \sn to fuse these
expressions. Our integration leverages TensorFlow's support for user-defined operators and graph
rewriting and makes no changes to the core TensorFlow engine.

\paragraph{NumPy and Pandas.}
\label{subsec:numpy-pandas-integration}

Our integrations with NumPy and Pandas required more effort because these libraries' APIs
are \emph{eagerly} evaluated.
To enable lazy evaluation in NumPy, we built a wrapper
for its array data type, \lstinline$ndarray$, which contains a \snobj. We also built
a \sn encoder to wrap the pointer to the data buffer in the
\lstinline$ndarray$ structure using the \sn vector type.\footnote{
NumPy's internal data format is already a packed array of primitive types.}
We overwrote the functions that print arrays
or extract elements from them to force evaluation.
Finally, all of our ported operators accept either a standard \lstinline$ndarray$ or our wrapper for their input arguments, and return a \snobj wrapper with those arguments as dependencies (\S\ref{sub:api}).


We ported Pandas in a similar way, by creating a wrapper object around dataframes.
We ported Pandas' filtering, sorting, predicate masking, aggregation, per-element
string slicing and \lstinline$getUniqueElements$ functions to \sn. 
Internally, Pandas represents dataframe columns as NumPy arrays, so we could use the same encoder as in NumPy and simply pass a pointer to these raw primitive arrays to \sn.

\section{Evaluation}
\label{sec:eval}

To evaluate \sn, we sought to answer the following questions:
(1) How much does \sn speed up data analytics workloads end-to-end, both within and across libraries?
(2) Does \sn exhibit performance benefits when deployed incrementally?
(3) Can \sn's generated code match the performance of existing optimized systems?

\paragraph{Experimental Setup.}
Unless otherwise noted, we ran tests on a server with an Intel Xeon E5-2680v3 CPU with 12 cores (24
hyperthreads) based on the Haswell micro-architecture and 128 GB of memory. Our C++ baselines are compiled using
LLVM 3.5 (with \code{O3} and \code{LTO}).  Results average over five runs.


\subsection{Accelerating Existing Frameworks}
\label{sub:eval-libraries}

\sn accelerates programs using individual libraries by applying optimizations across calls to the
library, fusing operators, and generating vectorized, multithreaded machine code. We benchmark \sn
integrations with the four frameworks from \S\ref{sec:integrations}: Spark 2.0, TensorFlow 0.12
(without the XLA compiler), TensorFlow 1.0 (with XLA ), NumPy 1.8.2, and Pandas 0.19.1.

\begin{figure}[t]
  \centering
  \begin{subfigure}[b]{0.49\columnwidth}
    \centering
    \includegraphics[width=35mm]{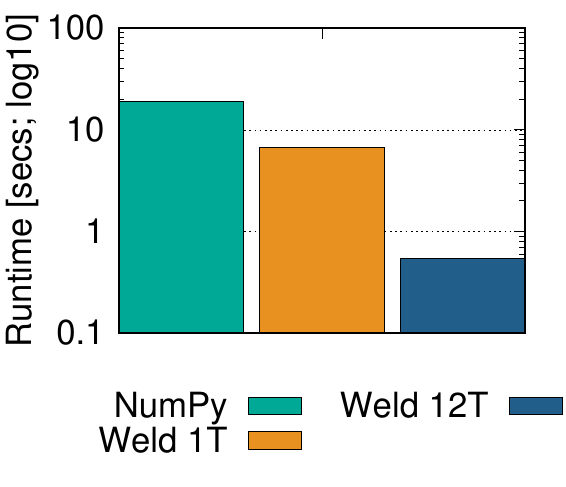}
    \caption{NumPy.}
    \label{fig:eval-libraries-numpy}
  \end{subfigure}
  \begin{subfigure}[b]{0.49\columnwidth}
    \centering
    \includegraphics[width=35mm]{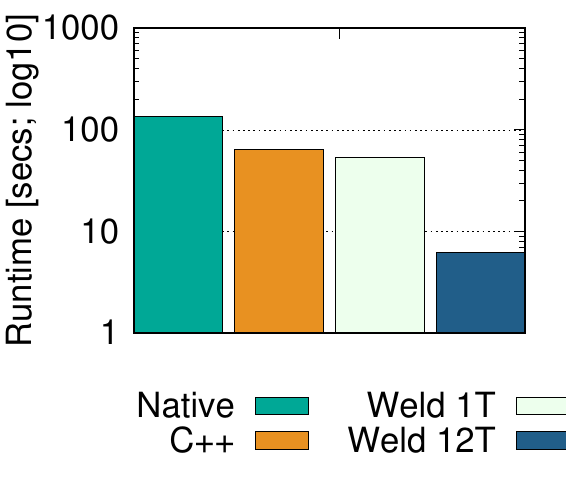}
    \caption{Pandas.}
    \label{fig:eval-libraries-pandas}
  \end{subfigure}
  \begin{subfigure}[b]{0.49\columnwidth}
    \centering
    \includegraphics[width=35mm]{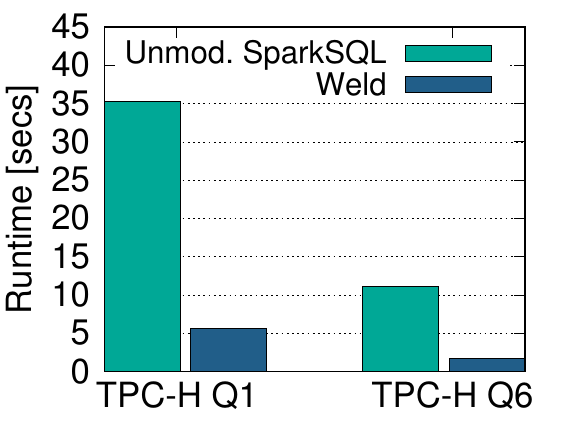}
    \caption{Spark SQL.}
    \label{fig:eval-libraries-sparksql}
  \end{subfigure}
  \begin{subfigure}[b]{0.49\columnwidth}
    \centering
    \includegraphics[width=35mm]{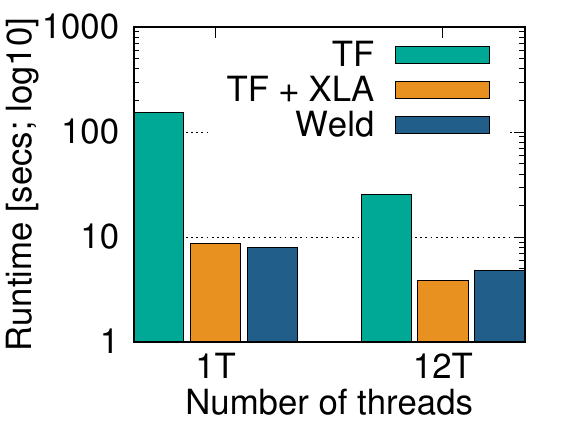}
    \caption{TensorFlow.}
    \label{fig:eval-libraries-tf-nn}
  \end{subfigure}
  \setlength{\belowcaptionskip}{-10pt}
  \caption{
    Workloads using individual libraries.
  }
\end{figure}

\paragraph{NumPy.}
We evaluate the NumPy \sn integration with a Black Scholes implementation on a set of 100 million
records. Black Scholes is a computationally expensive data parallel workload with mathematical
operators such as square roots, exponents, and logarithms.  Figure~\ref{fig:eval-libraries-numpy}
shows the results. \sn emits vectorized exponent and logarithm functions using AVX2, while NumPy
does not vectorize these calls. This leads to a 3$\times$ speedup over the native NumPy
implementation and a 33$\times$ improvement on 12 cores. \S\ref{sub:eval-ablation} breaks down the
speedups in this workload.

\paragraph{Pandas.}
We evaluate our Pandas integration on a data science workload from the Pandas
Cookbook~\cite{pandas-workload}, using a 6GB dataset of zipcodes. The workload uses Pandas to clean
the dataset by first ``slicing'' the zipcodes to represent each one with five digits, removing all
nonexistent zipcodes, and filtering duplicate zipcodes. The Pandas library we compare against
implements operators in either C or Cython already.

Figure~\ref{fig:eval-libraries-pandas} shows the results. \sn fuses each operator in Pandas; in this
data-intensive workload, materialization and redundant traversals over the data prove expensive.
This renders loop fusion useful, leading to a 2.5$\times$ speedup over native Pandas.  \sn also
facilitates multithreading and transparently parallelizes this workload (Pandas is a single-threaded
framework) without any application changes, enabling a 21.6$\times$ improvement over native Pandas
on 12 cores.

\paragraph{Spark SQL.}
To illustrate \sn's benefits in a distributed framework, we evaluate Spark SQL's \sn integration on
TPC-H queries 1 and 6 with 20 Amazon EC2 r3.xlarge worker instances (2 cores, 30 GB memory) and
800GB of TPC-H data (scale factor 800). Data was read from the Spark SQL in-memory cache. As shown
in Figure~\ref{fig:eval-libraries-sparksql}, \sn provides a 6.2$\times$ speedup for TPC-H Q1 and
6.5$\times$ for Q6. \sn's performance improvement comes largely from its ability to generate
low-level, vectorized machine code; Spark natively generates Java bytecode, which the Java JIT
compiler cannot vectorize.

\paragraph{TensorFlow.}
We evaluate TensorFlow by training a binary logistic regression classifier using the MNIST
dataset~\cite{mnist} to recognize a digit as zero or nonzero. We test the performance of the
default TensorFlow implementation against a \sn-enabled TensorFlow. We use two different versions of
TensorFlow: one with the XLA compiler~\cite{xla} and one without. XLA is a JIT compiler which
converts TensorFlow computations into machine code. The TensorFlow developers introduced XLA as an
optional feature in its 1.0 release after observing that data movement across TensorFlow operators
was a dominant cost~\cite{xla}; XLA performs many of the optimizations \sn does within TensorFlow
(\eg loop fusion).

Figure~\ref{fig:eval-libraries-tf-nn} shows the results. On a single core, \sn's lazy evaluation
allows \il fragments from each operator to be fused and optimized, demonstrating a 19.3$\times$
speedup over TensorFlow without XLA.  This speedup comes from loop fusion across operators enabled
by \sn's lazily evaluated runtime. \sn accelerates the computation by 1.09$\times$ over TensorFlow
with XLA enabled. Despite \sn's greater generality, its performance on this workload matches XLA, which is
specialized for TensorFlow's linear algebra routines. Unlike XLA, \sn's integration with TensorFlow
additionally enables optimization across other libraries as well, as shown in
\S\ref{sec:eval-pipelines}.

With 12 threads, the speedup over TensorFlow without XLA reduces to 5.3$\times$, since the \sn
version becomes bound on memory access; the overhead associated with TensorFlow's runtime also
becomes sizable. \sn is within 20\% of TensorFlow with XLA on 12 threads. The accuracy on a held-out
validation set is unchanged for all of our implementations.


\subsection{Optimizing Across Libraries}
\label{sec:eval-pipelines}

\begin{figure}[t]
  \begin{subfigure}[b]{0.49\columnwidth}
    \centering
    \includegraphics[width=35mm]{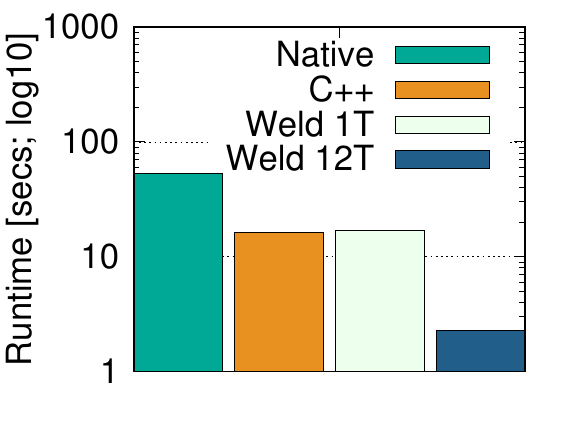}
    \caption{Pandas + NumPy, App. 1.}
    \label{fig:pandascomposition1}
  \end{subfigure}
  \begin{subfigure}[b]{0.49\columnwidth}
    \centering
    \includegraphics[width=35mm]{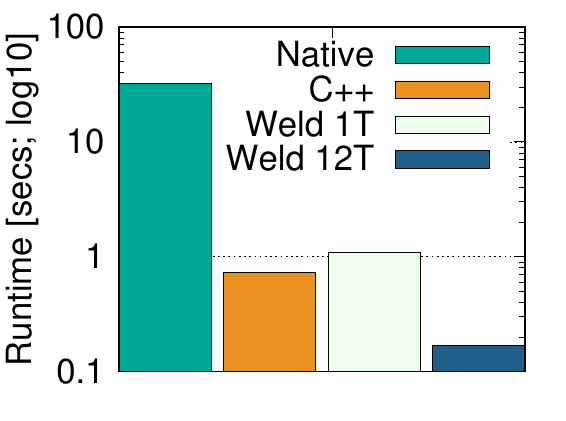}
    \caption{Pandas + NumPy, App. 2.}
    \label{fig:pandascomposition2}
  \end{subfigure}
  \begin{subfigure}[b]{0.49\columnwidth}
    \centering
    \includegraphics[width=35mm]{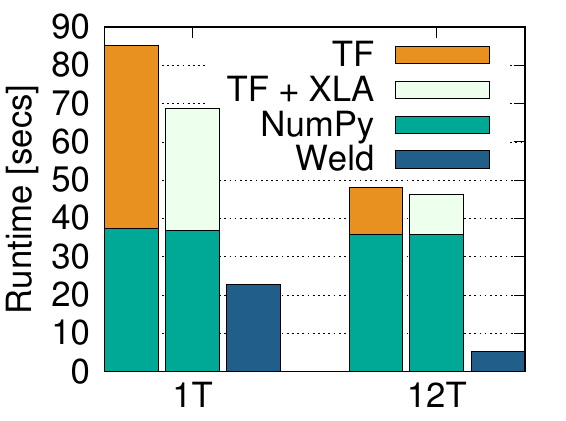}
    \caption{NumPy + TensorFlow.}
    \label{fig:numpytfcomposition}
  \end{subfigure}
  \begin{subfigure}[b]{0.49\columnwidth}
    \centering
    \includegraphics[width=35mm]{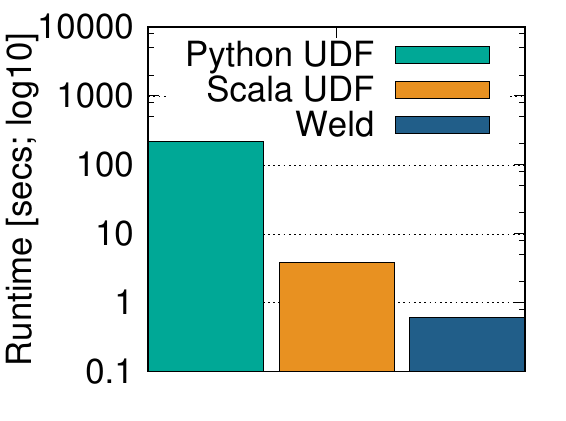}
    \caption{Spark SQL UDF.}
    \label{fig:eval-libraries-sparksql-udf}
  \end{subfigure}
  \setlength{\belowcaptionskip}{-10pt}
  \caption{
      Workloads which \emph{combine} libraries.
  }
\end{figure}

\sn also optimizes workflows which call into multiple libraries in a single application. We evaluate
our system on four representatitive workloads that combine libraries.

\paragraph{NumPy + Pandas.} To demonstrate the benefits of cross library optimization, we evaluate
\sn on workloads that combine NumPy and Pandas, adapted from the Pandas Cookbook~\cite{pandas-workload}.

In the first workload~\cite{pandas-workload}, we filter small cities in a population-by-cities dataframe, evaluate a
softmax regression model based on features in the dataframe, and aggregate the resulting crime
index predictions on a per-state basis. In the second workload, we perform the same filtering
operation then evaluate a linear model based on the same feature set, and finally compute an
average crime index across all cities. 

Figures~\ref{fig:pandascomposition1} and~\ref{fig:pandascomposition2} show the results. With \sn
integration, the API collects \il fragments \emph{across} library boundaries, enabling the NumPy
calls to be co-optimized with calls into Pandas. We observe speedups of up to 3.1$\times$ and
29$\times$ on a single core, and speedups of 23$\times$ and 187$\times$ across 12 cores.  These
optimizations come from cross-library loop fusion, vectorization of operators across libraries (\eg
using predication instructions from the Pandas call with arithmetic vector instructions generated
from the NumPy library call), and other standard whole-program optimizations such as inlining.
\S\ref{sub:eval-ablation} breaks down the speedups seen in the second workload in more detail.

\paragraph{NumPy + TensorFlow.} 
We applied \sn to a NumPy and TensorFlow image processing workflow. NumPy whitens images from
MNIST~\cite{lecun1998gradient} (a standard preprocessing task for image scoring) and TensorFlow
scores them using a standard logistic regression model trained to recognize digits.

Figure~\ref{fig:numpytfcomposition} shows the results.  With \sn integration we observe a 3$\times$
performance improvement over NumPy and TensorFlow with XLA, on a single thread. This performance
improvement increases to 8.9$\times$ over the native library implementations with 12 threads. \sn
provides a speedup despite TensorFlow's specialized XLA compiler optimizing the scoring computation
because it co-optimizes the image whitening task with the model scoring task.  With 12 cores, the
speedup is due to parallelizing the whitening computation; in the native library implementation of
the workload with 12 cores, TensorFlow parallelizes the model scoring but NumPy continues to run on
a single thread. Performance improvements over NumPy and TensorFlow without XLA were slightly
better. 

\paragraph{Spark SQL UDF.} We also demonstrate the benefits of operator composition by comparing a
\sn-enabled UDF with Scala and Python UDFs in Spark SQL.  The query is similar to one of our Pandas
workloads: it evaluates a linear model implemented by the UDF on each row of a 2.4GB dataset and
computes the average result.  The \sn-enabled version of the UDF was written in Python and
translated to \sn automatically using the library described in \S~\ref{sub:pyudfs}. 

Figure~\ref{fig:eval-libraries-sparksql-udf} shows the results. Calling the Scala and Python UDFs
for each row is expensive due to data marshaling between Spark SQL and these languages. In contrast,
the \sn UDF is co-optimized with the rest of the query (implemented using \sn-enabled Spark SQL
operators) and the program is vectorized, producing a 6$\times$ speedup over the Scala UDF and a
360$\times$ speedup over the Python UDF. We note here that despite the performance cost of UDFs, in
practice many analytics jobs use them~\cite{sparksql,shark}. This indicates that developers seem
unwilling to invest time in crafting fast parallel code in these use cases which increases the value
of a runtime that transparently accelerates workflows.


\subsection{Incremental Integration}
\label{sub:eval-incremental}

\begin{figure}[t]
  \begin{subfigure}[b]{0.49\columnwidth}
    \centering
    \includegraphics[width=35mm]{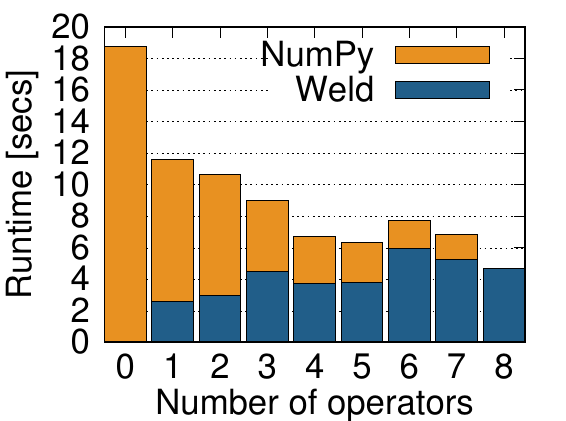}
    \caption{Black Scholes 1T.}
    \label{fig:incremental-st}
  \end{subfigure}
  \begin{subfigure}[b]{0.49\columnwidth}
    \centering
    \includegraphics[width=35mm]{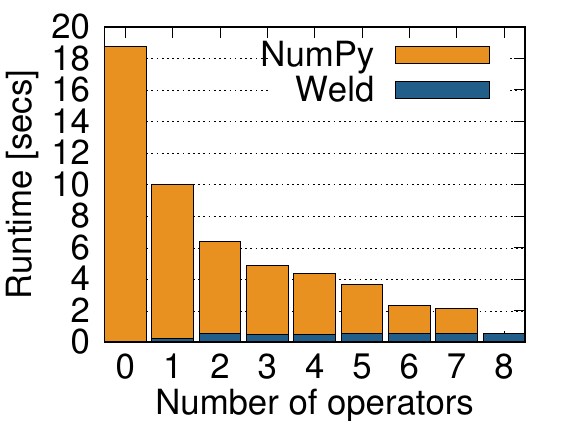}
    \caption{Black Scholes 12T.}
    \label{fig:incremental-mt}
  \end{subfigure}
  \setlength{\belowcaptionskip}{-10pt}
  \caption{
      Incremental Integration with one and 12 threads.}
    \label{fig:incremental-total}
\end{figure}

To show that \sn is incrementally deployable, we ran the Black Scholes workload from
\S~\ref{sec:eval-pipelines} on a single thread and incrementally implemented one operator at a time
using \sn's interface. Operators which were not \sn-enabled used native NumPy.  The Black Scholes
workload uses eight unique operators; we measured which operators were the most expensive in terms
of total number of CPU cycles and ported them from most to least computationally expensive.

Figure~\ref{fig:incremental-total} shows the results. On a single core, implementing the first
operator in \sn (the erf function, which evaluates an exponent and a closed integral) gives a
1.6$\times$ speedup, and implementing half the operators gives a 2.7$\times$ speedup over native
NumPy, by providing vectorized implementations for functions which NumPy runs sequentially.
Implementing more operators shows further (yet diminishing) speedups. With 12 threads, the single
threaded NumPy operators prove to be a bottleneck, a consequence of Amdahl's Law. Implementing half
the operators in \sn gives a modest 4$\times$ speedup, implementing all but one operator in \sn
gives a 9$\times$ speedup, and implementing all the operators in \sn gives a 33$\times$ speedup
since the entire workflow can be parallelized.


\subsection{CPU Backend Performance}
\label{sub:cpu-eval-ubench}

We now evaluate \sn's code generation on x86 CPUs by comparing the performance of a handful of data
processing workloads against several state-of-the-art, domain-specific compilers and systems. 

\begin{figure}[t]
  \centering
  \begin{subfigure}[b]{0.32\columnwidth}
    \centering
    \includegraphics[width=23mm]{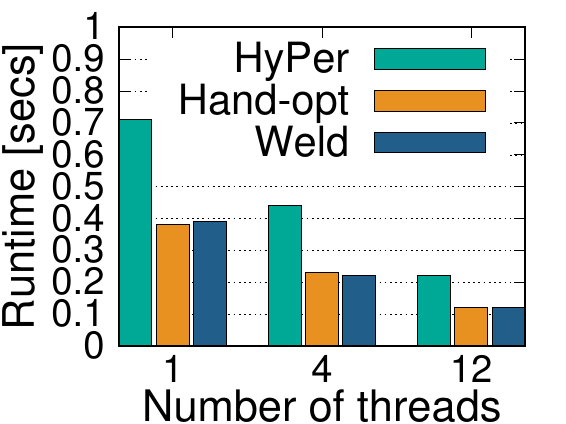}
    \caption{TPC-H Q1.}
    \label{fig:eval-tpch-q1}
  \end{subfigure}
  \begin{subfigure}[b]{0.32\columnwidth}
    \centering
    \includegraphics[width=23mm]{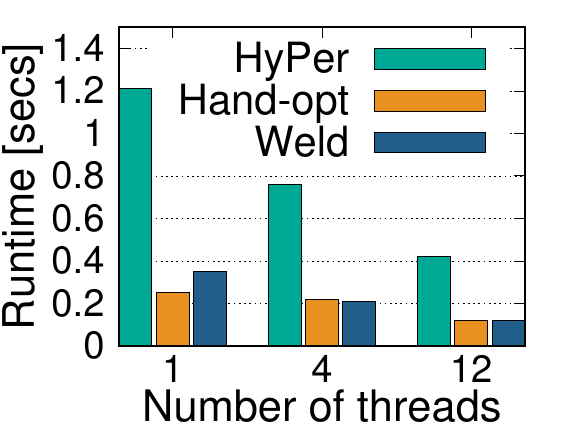}
    \caption{TPC-H Q3.}
    \label{fig:eval-tpch-q3}
  \end{subfigure}
  \begin{subfigure}[b]{0.32\columnwidth}
    \centering
    \includegraphics[width=23mm]{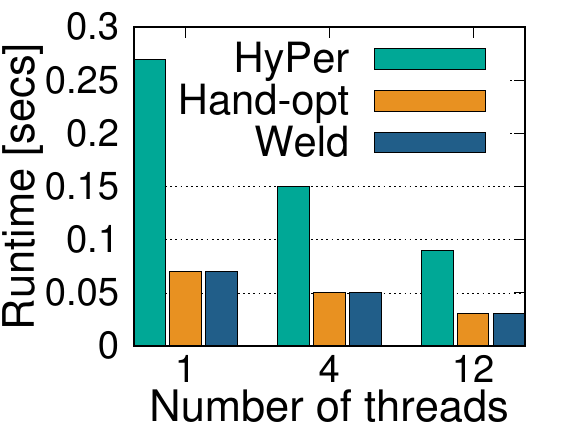}
    \caption{TPC-H Q6.}
    \label{fig:eval-tpch-q6}
  \end{subfigure}
  \begin{subfigure}[b]{0.32\columnwidth}
    \centering
    \includegraphics[width=23mm]{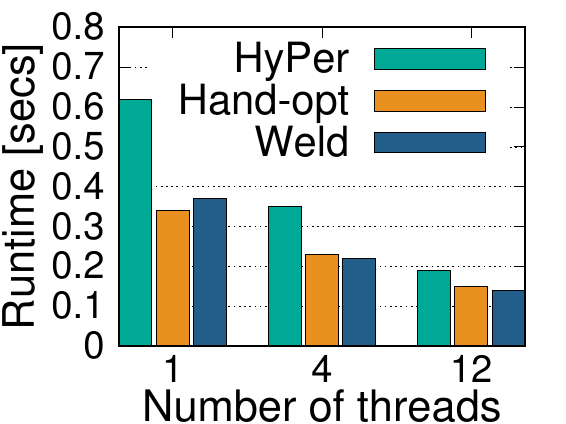}
    \caption{TPC-H Q12.}
    \label{fig:eval-tpch-q12}
  \end{subfigure}
  \begin{subfigure}[b]{0.32\columnwidth}
    \centering
    \includegraphics[width=23mm]{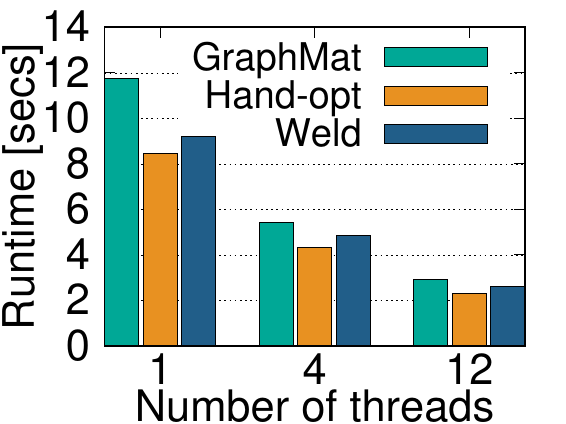}
    \caption{PageRank.}
    \label{fig:eval-pagerank}
  \end{subfigure}
  \setlength{\belowcaptionskip}{-10pt}
  \caption{CPU microbenchmarks on TPC-H and PageRank}
  \label{fig:eval-tpch}
\end{figure}

\paragraph{TPC-H queries.}
Figure~\ref{fig:eval-tpch} show the results for the TPC-H queries (scale factor 10), compared
against the HyPer~\cite{hyper} database and a hand-optimized C baseline using Intel's AVX2
intrinsics for vectorization and OpenMP~\cite{openmp} for parallelization. HyPer generates LLVM \il
for SQL queries, which is then compiled to machine code before execution. \sn uses the same query
plan as HyPer; query plans were obtained from~\cite{hyperplans}.  Execution time is competitive with
HyPer across the board.  \sn outperforms HyPer on Q6 and Q12 because it applies predication in the
\il and generates explicitly vectorized LLVM \il. HyPer depends on LLVM for vectorization, which
does not automatically perform predication.

\eat {
\paragraph{Nearest Neighbor Classification.}
Figure~\ref{fig:eval-nn} shows the results of running a nearest neighbor classifier acquired from a
set of TensorFlow applications~\cite{nngithub} with and without \sn integration, and against a
handwritten Eigen~\cite{eigen} implementation.  \sn's generated code outperforms TensorFlow by
4.64$\times$ on a single thread and by 5.06$\times$ on 12 threads; \sn's performance is competitive
with the Eigen program.
}

\paragraph{Linear Algebra.} Figure~\ref{fig:eval-libraries-tf-nn} shows a comparison of \sn
against TensorFlow's specialized XLA compiler. As we previously reported, \sn's generated code
matches the performance of the XLA's generated code, despite \sn optimizing a more general \il (XLA
is specialized for TensorFlow's linear algebra operators).

\paragraph{PageRank.} Figure~\ref{fig:eval-pagerank} shows results for a PageRank implementation
written in the \sn \il, compared against the GraphMat~\cite{graphmat} graph processing framework.
GraphMat had the fastest multicore PageRank implementation we found. \sn's per-iteration runtime for
both serial and parallel code outperforms GraphMat, and is competitive with a hand-optimized C
baseline which uses Cilk~\cite{cilk} for multi-threading.

In all cases, \sn's runtime memory usage also matched the memory usage of our C implementations.  In
summary, \sn produces machine code competitive with existing systems on a variety of workloads,
demonstrating both the expressiveness of the \il and the effectiveness of our optimizer.


\subsection{GPU Backend Performance}
\label{sub:gpu-eval-ubench}

\begin{figure}[t]
  \centering
  \includegraphics[width=80mm]{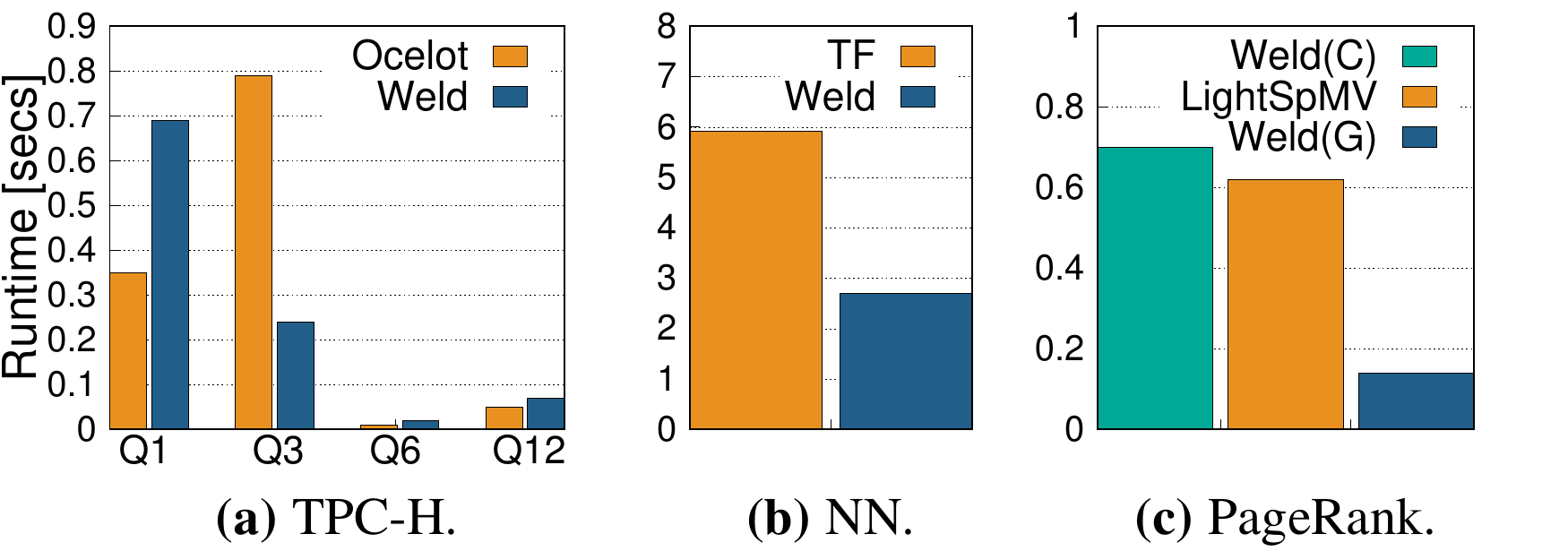}
  \caption{GPU microbenchmarks on TPC-H, PageRank, and nearest neighbor classification.}
  \label{fig:eval-gpu-tpch-nn-pagerank}
  \setlength{\belowcaptionskip}{-10pt}
\end{figure}

\sn's \il is designed to support diverse hardware platforms. We ran a set of \sn programs on a
prototype GPU backend to illustrate that the abstractions in the \il are sufficient to enable fast
code generation across platforms. We compare our GPU results to other existing systems which also
target GPU execution using an Nvidia GeForce GTX Titan X with 12 GB of internal DDR5 memory running
Nvidia's OpenCL implementation (version 367.57, CUDA 7.5).

\paragraph{TPC-H queries.}

Figure~\ref{fig:eval-gpu-tpch-nn-pagerank}a shows the results of running the same four TPC-H queries
from~~\S\ref{sub:cpu-eval-ubench} against Ocelot~\cite{ocelot}, a database optimized for GPUs. We
find that for most queries, \sn generates OpenCL code that generally outperforms the CPU backend and
is on par with Ocelot, apart from Q1 where it is within a factor of 2 of Ocelot.

\paragraph{Nearest Neighbor Classification.}

Figure~\ref{fig:eval-gpu-tpch-nn-pagerank}b shows the results of running the same nearest neighbor
classifier as before, using TensorFlow (with its GPU backend enabled) with
and without \sn integration. \sn's generated code outperforms TensorFlow's
handwritten CUDA operator by 2.2$\times$.

\paragraph{PageRank.}

The PageRank implementation is based on the un-nesting strategy described in
\S~\ref{sec:implementation} without any dynamic load balancing. Nevertheless, the GPU-based
implementation achieves a roughly 5$\times$ speedup over the CPU-based implementation on account of
the high degree of data parallelism that can be exploited by the GPU. For reference,
Figure~\ref{fig:eval-gpu-tpch-nn-pagerank}c includes the per-iteration performance of the LightSpMV
implementation of PageRank~\cite{liu2015lightspmv} which is, to the best of our knowledge, the
fastest implementation on the compressed sparse row representation of the graph. LightSpMV, however,
implements dynamic load balancing instead of static partitioning which comes with a $4\times$
runtime overhead.


\subsection{Effects of Individual Optimizations}
\label{sub:eval-ablation}

\begin{figure}[t]
  \begin{subfigure}[b]{0.49\columnwidth}
    \centering
    \includegraphics[width=35mm]{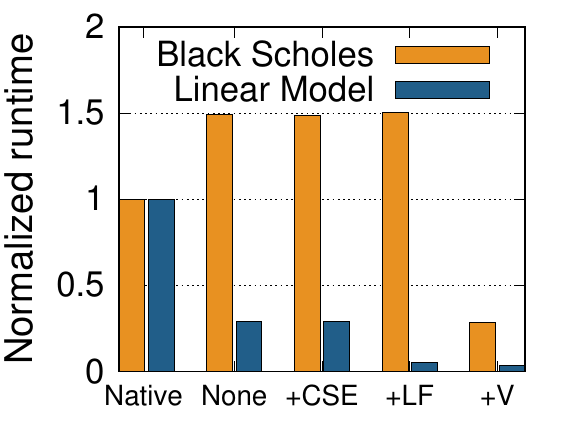}
    \caption{Adding Optimizations}
    \label{fig:ablation-add}
  \end{subfigure}
  \begin{subfigure}[b]{0.49\columnwidth}
    \centering
    \includegraphics[width=35mm]{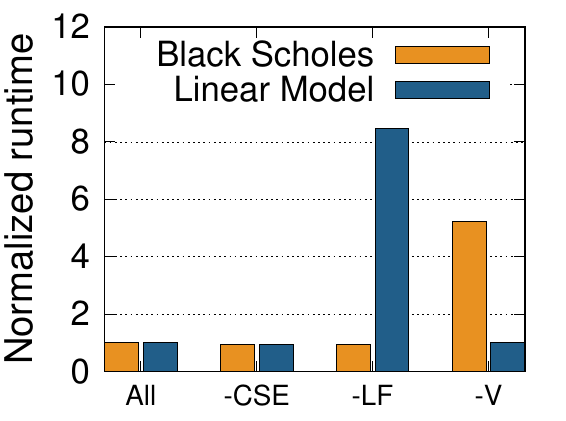}
    \caption{Removing Optimizations}
    \label{fig:ablation-remove}
  \end{subfigure}
  \setlength{\belowcaptionskip}{-10pt}
  \caption{Effects of individual optimizations on Black Scholes and the Pandas linear model
    workload.  CSE, LF, and V correspond to Common Subexpression Elimination, Loop Fusion, and
    Vectorization respectively.  }
  \label{fig:ablation}
\end{figure}

We evaluate the effect individual optimizations have on two tasks introduced earlier in the
evaluation section: the Black Scholes workload that uses NumPy and the linear model workload that
uses Pandas and NumPy. For each experiment, we show the effect of incrementally adding optimizations
in \sn and the effect of removing single optimizations from the entire optimization suite.
Figure~\ref{fig:ablation} shows the results obtained by running both workloads on a single thread.

The Black Scholes and Pandas workloads differ in that the Black Scholes workload is \emph{compute
intensive} and does not benefit from data movement optimizations. This is evident in the results;
the only transformation which has a substantial impact is vectorization, which makes more effective use
of the execution units on a single core. In contrast, the Pandas workload is \emph{data intensive},
with only a few cycles of computation per byte; most time goes into performing scans over memory.
Loop fusion provides the greatest gain here by transforming the entire computation to take a single
traversal over the data and preventing intermediate results from being written back to memory.


\subsection{Builder Abstraction}
\label{sub:eval-builders}

\begin{figure}[t]
  \centering
  \includegraphics[width=80mm]{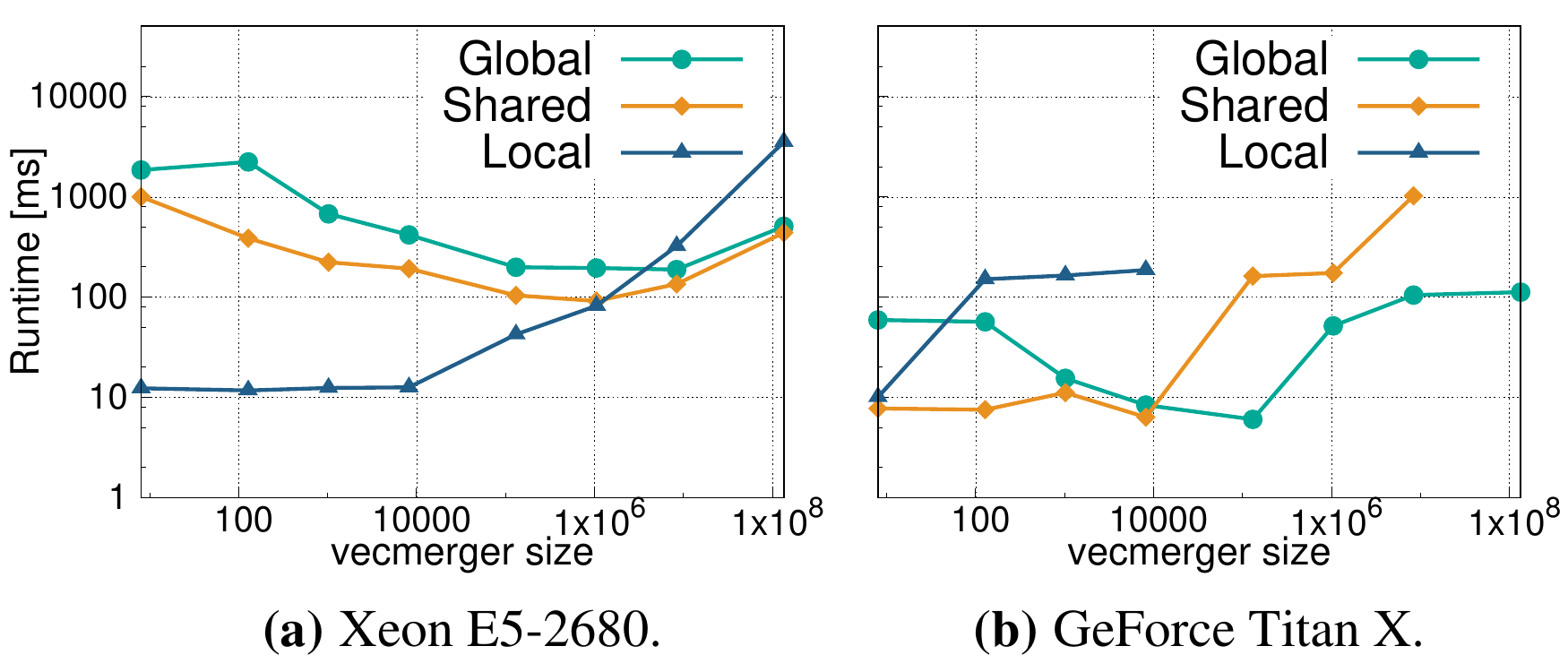}
  \caption{Builder implementations on heterogeneous hardware. Missing data
    points indicate that the strategy ran out of memory or did not finish.}
  \label{fig:builder-hardware}
  \setlength{\belowcaptionskip}{-10pt}
  \squeezeup
\end{figure}

As described in \S\ref{sec:implementation}, builders in \sn are implemented differently on CPUs and GPUs.
In this section we further motivate
why builders are a powerful abstraction by considering several potential implementation strategies
for the \lstinline$vecmerger$ builder on the CPU and GPU. The \lstinline$vecmerger$
takes an input vector and allows merges of new values into arbitrary positions in the vector. We
use the \lstinline$vecmerger$ to implement a \sn program that counts the number of occurrences of
each key in a list.
We tested three different implementation strategies for the \lstinline$vecmerger$ on each platform
-- thread-local copies combined with a final aggregation step (local), thread block- or NUMA
node-local copies updated with atomic instructions and combined with a smaller final aggregation
(shared), and a single global copy updated with atomic instructions (global). Each strategy ran with
all available threads on the platform. All runs merged $10^8$ 32-bit integer keys.

Figure~\ref{fig:builder-hardware} shows that the size at which each
\lstinline$vecmerger$ strategy becomes optimal varies widely across the two platforms.
Even if a developer wrote code in a generic hardware-independent language, such as OpenCL~\cite{opencl},
to choose the strategy dynamically, this code would not be portable across platforms due to
different transition points.
In contrast, a runtime based on the builder abstraction, such as \sn, can generate completely different
code for each platform to optimize the \lstinline$vecmerger$ on it.
(Note that we have not yet implemented all the variants here in our prototype.)

\subsection{Compilation Times}

\sn's compile times (including optimization of the \il and code generation via LLVM) ranged from 62 ms to
257 ms (mean 126 ms and median 117 ms) across all experiments. Since we expect most analytics
workloads on large datasets to run for several seconds, we believe that these times are acceptable
for \sn's target applications.


\section{Related Work}
\label{sec:related}

\lstset{language=NVL}

\sn builds on ideas in multiple fields, including compilers, parallel programming models and database engines.
Unlike existing systems, however, \sn aims to design an interface that can be used to optimize across \emph{diverse existing libraries} instead of creating a monolithic programming model in which all applications should be built.
We show that such cross-library optimization is crucial for performance in workloads that combine multiple libraries.
In addition, unlike systems that do runtime code generation for more specific workloads (\eg databases~\cite{hyper} or linear algebra~\cite{xla}), \sn offers a small and general \il that can achieve state-of-the-art performance \emph{across} these workloads.

Multiple prior systems aim to simplify programming parallel hardware.
Hardware-independent intermediate languages such as OpenCL~\cite{opencl}, LLVM~\cite{llvm}, and SPIR~\cite{spir} are based on low-level sequential instruction sets where threads communicate via shared memory, which makes it hard to perform complex optimizations for \emph{parallel} code, such as loop fusion or loop tiling across parallel functions.
Moreover, the APIs to these systems are evaluated eagerly, resulting in multiple independent invocations when applications combine multiple libraries.
In~.NET languages, LINQ~\cite{linq} offers a lazily evaluated API that has been used to target clusters~\cite{dryadlinq} and heterogeneous hardware~\cite{dandelion} using relational algebra as an \il.
However, LINQ is designed for the ``closed'' world of .NET programs and does not provide a means of interfacing with code outside the .NET VM.
Moreover, although its relational \il can support various fusion optimizations~\cite{steno}, it is difficult to express other optimizations such as loop tiling.
Spark~\cite{cacm-spark}, FlumeJava~\cite{flumejava} and other systems~\cite{musketeer,emma} also perform optimizations using a relational or functional \il, while
Pydron~\cite{pydron} uses annotations to parallelize Python code.

Runtime code generation has been used in systems including databases~\cite{hyper,sparksql,tupleware} and TensorFlow's XLA compiler~\cite{xla}.
However, most of these systems are limited to a narrow workload domain (\eg linear algebra or SQL with sequential user-defined functions~\cite{tupleware}).
Our work shows that \sn's more general \il, together with its optimizer, enables code generation on par with these systems for multiple key workloads and simultaneously allows optimizing across them.

The compilers literature is also rich with parallel languages, \ils and domain-specific languages (DSLs)~\cite{tapir,cilk,dphaskell,nesl,delite}.
However, most of this work focuses on static compilation of an entire program, whereas \sn allows dynamic compilation at runtime, even when programs load libraries dynamically or compose them in a dynamically typed language like Python.
\sn's \il is closest to monad comprehensions~\cite{monad-comprehensions} and to
Delite's DMLL~\cite{delite-popl,dmll}, both of which support
nested parallel loops that can update multiple results per iteration.\footnote{
\sn has some differences from these \ils, however, most notably that it represents
builders as a first-class type. For example, in Delite, the result of each loop is a
collection, not a builder, meaning that nested loops cannot efficiently update the same
builder unless the compiler treats this case specially.
}
Builders are also similar to Cilk reducers~\cite{cilk-reducers} and to LVars~\cite{lvars},
although these systems to not implement them in different ways on different hardware platforms.
Finally, \sn's optimization techniques (\eg loop fusion) are standard, but our contribution is to demonstrate that they can be applied easily to \sn's \il and yield high-quality code even when libraries are combined using the narrow interface \sn provides.


\section{Conclusion}
\label{sec:conclusions}

We presented \sn, a novel interface and runtime for accelerating data-intensive workloads using
disjoint functions and libraries. Today, these workloads often perform an order of magnitude below
hardware limits due to extensive data movement. \sn addresses this problem through an intermediate
representation (\il) that can capture data-parallel applications and a runtime API that allows \il
code from different libraries to be combined. Together, these ideas allow \sn to bring powerful
optimizations to multi-library workflows without changing their user-facing APIs.  We showed that
\sn is easy to integrate into Spark SQL, TensorFlow, Pandas and NumPy and provides speedups of up to
$6.5\times$ in workflows using a single library and 29$\times$ across libraries.

\section{Acknowledgements}
\label{sec:acknowledgements}

We thank Pratiksha Thaker, Chirstopher Aberger, Firas Abuzaid, Parimarjan Negi,
Paroma Varma, Peter Bailis, and the many members of the Stanford InfoLab for
their valuable feedback on this work. This research was supported in part by
affiliate members and other supporters of the Stanford DAWN project -- Intel,
Microsoft, Teradata, and VMware -- as well as NSF CAREER grant CNS-1651570 and
NSF Graduate Research Fellowship under grant DGE-1656518.  Any opinions,
findings, and conclusions or recommendations expressed in this material are
those of the author(s) and do not necessarily reflect the views of the National
Science Foundation.

\footnotesize \bibliographystyle{IEEEtran}
\bibliography{nvl}

\begin{thebibliography}{10}
\providecommand{\url}[1]{#1}
\csname url@samestyle\endcsname
\providecommand{\newblock}{\relax}
\providecommand{\bibinfo}[2]{#2}
\providecommand{\BIBentrySTDinterwordspacing}{\spaceskip=0pt\relax}
\providecommand{\BIBentryALTinterwordstretchfactor}{4}
\providecommand{\BIBentryALTinterwordspacing}{\spaceskip=\fontdimen2\font plus
\BIBentryALTinterwordstretchfactor\fontdimen3\font minus
  \fontdimen4\font\relax}
\providecommand{\BIBforeignlanguage}[2]{{%
\expandafter\ifx\csname l@#1\endcsname\relax
\typeout{** WARNING: IEEEtran.bst: No hyphenation pattern has been}%
\typeout{** loaded for the language `#1'. Using the pattern for}%
\typeout{** the default language instead.}%
\else
\language=\csname l@#1\endcsname
\fi
#2}}
\providecommand{\BIBdecl}{\relax}
\BIBdecl

\bibitem{sparksql}
M.~Armbrust, R.~S. Xin, C.~Lian, Y.~Huai, D.~Liu, J.~K. Bradley, X.~Meng,
  T.~Kaftan, M.~J. Franklin, A.~Ghodsi, and M.~Zaharia, ``{Spark SQL:
  Relational Data Processing in Spark},'' in \emph{Proc. ACM SIGMOD}, 2015, pp.
  1383--1394.

\bibitem{numpy}
\BIBentryALTinterwordspacing
{NumPy}. [Online]. Available: \url{http://www.numpy.org/}
\BIBentrySTDinterwordspacing

\bibitem{pandas}
W.~McKinney, ``{ Data Structures for Statistical Computing in Python },'' in
  \emph{Proceedings of the 9th Python in Science Conference}, 2010, pp. 51 --
  56.

\bibitem{tensorflow}
M.~Abadi, P.~Barham, J.~Chen, Z.~Chen, A.~Davis, J.~Dean, M.~Devin,
  S.~Ghemawat, G.~Irving, M.~Isard \emph{et~al.}, ``{TensorFlow: A System for
  Large-Scale Machine Learning},'' in \emph{Proc. {USENIX OSDI}}, 2016, pp.
  265--283.

\bibitem{kagi1996memory}
A.~Kagi, J.~R. Goodman, and D.~Burger, ``{Memory Bandwidth Limitations of
  Future Microprocessors},'' in \emph{ISCA}, 1996, pp. 78--78.

\bibitem{blas}
C.~L. Lawson, R.~J. Hanson, D.~R. Kincaid, and F.~T. Krogh, ``{Basic Linear
  Algebra Subprograms for Fortran Usage},'' \emph{ACM Trans. Math. Softw.},
  vol.~5, no.~3, pp. 308--323, 1979.

\bibitem{semcache}
\BIBentryALTinterwordspacing
N.~AlSaber and M.~Kulkarni, ``Semcache: Semantics-aware caching for efficient
  gpu offloading,'' in \emph{Proceedings of the 27th International ACM
  Conference on International Conference on Supercomputing}, ser. ICS '13,
  2013. [Online]. Available: \url{http://doi.acm.org/10.1145/2464996.2465021}
\BIBentrySTDinterwordspacing

\bibitem{loop-tiling}
\BIBentryALTinterwordspacing
Loop tiling. [Online]. Available:
  \url{https://en.wikipedia.org/wiki/Loop\_tiling}
\BIBentrySTDinterwordspacing

\bibitem{opencl}
J.~E. Stone, D.~Gohara, and G.~Shi, ``{OpenCL: A Parallel Programming Standard
  for Heterogeneous Computing Systems},'' \emph{Computing in Science
  Engineering}, vol.~12, no.~3, pp. 66--73, 2010.

\bibitem{cuda}
\BIBentryALTinterwordspacing
{CUDA}. [Online]. Available:
  \url{http://www.nvidia.com/object/cuda\_home\_new.html}
\BIBentrySTDinterwordspacing

\bibitem{hyper}
T.~Neumann, ``{Efficiently Compiling Efficient Query Plans for Modern
  Hardware},'' \emph{Proc. {VLDB}}, vol.~4, no.~9, pp. 539--550, 2011.

\bibitem{xla}
\BIBentryALTinterwordspacing
{TensorFlow XLA}. [Online]. Available:
  \url{https://www.tensorflow.org/performance/xla/}
\BIBentrySTDinterwordspacing

\bibitem{spark-wholestage}
\BIBentryALTinterwordspacing
S.~Agarwal, D.~Liu, and R.~Xin, ``{Apache Spark} as a compiler: Joining a
  billion rows per second on a laptop.'' [Online]. Available:
  \url{https://databricks.com/blog/2016/05/23/}
\BIBentrySTDinterwordspacing

\bibitem{pandas-workload}
\BIBentryALTinterwordspacing
{Pandas Cookbook example}. [Online]. Available:
  \url{http://nbviewer.jupyter.org/github/jvns/pandas-cookbook/blob/v0.1/cookbook/Chapter%207%20-%20Cleaning%20up%20messy%20data.ipynb}
\BIBentrySTDinterwordspacing

\bibitem{ousterhout2015}
K.~Ousterhout, R.~Rasti, S.~Ratnasamy, S.~Shenker, and B.-G. Chun, ``{Making
  Sense of Performance in Data Analytics Frameworks},'' in \emph{12th USENIX
  Symposium on Networked Systems Design and Implementation (NSDI 15)}, 2015,
  pp. 293--307.

\bibitem{tupleware}
A.~Crotty, A.~Galakatos, K.~Dursun, T.~Kraska, C.~Binnig, U.~Cetintemel, and
  S.~Zdonik, ``{An Architecture for Compiling UDF-centric Workflows},''
  \emph{Proc. VLDB Endow.}, vol.~8, no.~12, pp. 1466--1477, August 2015.

\bibitem{hogwild}
B.~Recht, C.~Re, S.~Wright, and F.~Niu, ``{Hogwild: A Lock-Free Approach to
  Parallelizing Stochastic Gradient Descent},'' in \emph{Advances in Neural
  Information Processing Systems 24}, J.~Shawe-Taylor, R.~S. Zemel, P.~L.
  Bartlett, F.~Pereira, and K.~Q. Weinberger, Eds.\hskip 1em plus 0.5em minus
  0.4em\relax Curran Associates, Inc., 2011, pp. 693--701.

\bibitem{linq}
E.~Meijer, B.~Beckman, and G.~Bierman, ``{{LINQ}: Reconciling Object, Relations
  and XML in the .NET Framework},'' in \emph{SIGMOD}.\hskip 1em plus 0.5em
  minus 0.4em\relax ACM, 2006, pp. 706--706.

\bibitem{monad-comprehensions}
T.~Grust, \emph{Monad Comprehensions: A Versatile Representation for
  Queries}.\hskip 1em plus 0.5em minus 0.4em\relax Berlin, Heidelberg: Springer
  Berlin Heidelberg, 2004, pp. 288--311.

\bibitem{walker-types}
D.~Walker, ``{Substructural Type Systems},'' in \emph{Advanced Topics in Types
  and Programming Languages}, B.~C. Pierce, Ed.\hskip 1em plus 0.5em minus
  0.4em\relax MIT Press, 2005, ch.~1.

\bibitem{cacm-spark}
M.~Zaharia, R.~S. Xin, P.~Wendell, T.~Das, M.~Armbrust, A.~Dave, X.~Meng,
  J.~Rosen, S.~Venkataraman, M.~J. Franklin, A.~Ghodsi, J.~Gonzalez,
  S.~Shenker, and I.~Stoica, ``{Apache Spark: A Unified Engine for Big Data
  Processing},'' \emph{Commun. ACM}, vol.~59, no.~11, pp. 56--65, October 2016.

\bibitem{dryadlinq}
Y.~Yu, M.~Isard, D.~Fetterly, M.~Budiu, U.~Erlingsson, P.~K. Gunda, and
  J.~Currey, ``{DryadLINQ: A System for General-purpose Distributed
  Data-parallel Computing Using a High-level Language},'' in \emph{Proceedings
  of the 8th USENIX Conference on Operating Systems Design and Implementation},
  ser. OSDI'08.\hskip 1em plus 0.5em minus 0.4em\relax Berkeley, CA, USA:
  USENIX Association, 2008, pp. 1--14.

\bibitem{dandelion}
C.~J. Rossbach, Y.~Yu, J.~Currey, J.-P. Martin, and D.~Fetterly, ``{Dandelion:
  A Compiler and Runtime for Heterogeneous Systems},'' in \emph{Proc. ACM
  SOSP}.\hskip 1em plus 0.5em minus 0.4em\relax ACM, 2013, pp. 49--68.

\bibitem{pydron}
S.~C. M{\"u}ller, G.~Alonso, A.~Amara, and A.~Csillaghy, ``{Pydron:
  Semi-Automatic Parallelization for Multi-Core and the Cloud.}'' in
  \emph{OSDI}, 2014, pp. 645--659.

\bibitem{llvm}
C.~Lattner and V.~Adve, ``{LLVM: a compilation framework for lifelong program
  analysis transformation},'' in \emph{Code Generation and Optimization, 2004.
  CGO 2004. International Symposium on}, 2004, pp. 75--86.

\bibitem{avx2}
\BIBentryALTinterwordspacing
{Intel AVX2}. [Online]. Available:
  \url{https://software.intel.com/en-us/node/523876}
\BIBentrySTDinterwordspacing

\bibitem{cilk}
R.~D. Blumenofe, C.~F. Joerg, B.~C. Kurzmaul, C.~E. Leiserson, K.~H. Randall,
  and Y.~Zhou, ``{Cilk: An Efficient Multithreaded Runtime System},''
  \emph{Journal of Parallel and Distributed Computing}, vol.~37, no.~1, pp.
  55--69, 1996.

\bibitem{blelloch1988compiling}
G.~E. Blelloch and G.~W. Sabot, ``{Compiling Collection-oriented Languages onto
  Massively Parallel Computers},'' in \emph{Frontiers of Massively Parallel
  Computation, 1988. Proceedings., 2nd Symposium on the Frontiers of}.\hskip
  1em plus 0.5em minus 0.4em\relax IEEE, 1988, pp. 575--585.

\bibitem{mnist}
\BIBentryALTinterwordspacing
{MNIST}. [Online]. Available: \url{http://yann.lecun.com/exdb/mnist/}
\BIBentrySTDinterwordspacing

\bibitem{lecun1998gradient}
Y.~LeCun, L.~Bottou, Y.~Bengio, and P.~Haffner, ``{Gradient-based Learning
  Applied to Document Recognition},'' \emph{Proceedings of the IEEE}, vol.~86,
  no.~11, pp. 2278--2324, 1998.

\bibitem{shark}
R.~S. Xin, J.~Rosen, M.~Zaharia, M.~J. Franklin, S.~Shenker, and I.~Stoica,
  ``{Shark: SQL and Rich Analytics at Scale},'' in \emph{Proceedings of the
  2013 ACM SIGMOD International Conference on Management of Data}, ser. SIGMOD
  '13.\hskip 1em plus 0.5em minus 0.4em\relax ACM, 2013, pp. 13--24.

\bibitem{openmp}
\BIBentryALTinterwordspacing
{OpenMP}. [Online]. Available: \url{http://openmp.org/wp/}
\BIBentrySTDinterwordspacing

\bibitem{hyperplans}
\BIBentryALTinterwordspacing
{HyPer Web Interface}. [Online]. Available:
  \url{http://hyper-db.de/interface.html}
\BIBentrySTDinterwordspacing

\bibitem{graphmat}
N.~Sundaram, N.~Satish, M.~M.~A. Patwary, S.~R. Dulloor, M.~J. Anderson, S.~G.
  Vadlamudi, D.~Das, and P.~Dubey, ``{GraphMat: High Performance Graph
  Analytics Made Productive},'' \emph{Proc. {VLDB}}, vol.~8, no.~11, pp.
  1214--1225, 2015.

\bibitem{ocelot}
M.~Heimel, M.~Saecker, H.~Pirk, S.~Manegold, and V.~Markl,
  ``{Hardware-oblivious parallelism for in-memory column-stores},''
  \emph{Proceedings of the VLDB Endowment}, vol.~6, no.~9, pp. 709--720, 2013.

\bibitem{liu2015lightspmv}
Y.~Liu and B.~Schmidt, ``{LightSpMV: faster CSR-based sparse matrix-vector
  multiplication on CUDA-enabled GPUs},'' in \emph{2015 IEEE 26th International
  Conference on Application-specific Systems, Architectures and Processors
  (ASAP)}.\hskip 1em plus 0.5em minus 0.4em\relax IEEE, 2015, pp. 82--89.

\bibitem{spir}
\BIBentryALTinterwordspacing
J.~Kessenich. (2015) An introduction to {SPIR-V}. [Online]. Available:
  \url{https://www.khronos.org/registry/spir-v/papers/WhitePaper.pdf}
\BIBentrySTDinterwordspacing

\bibitem{steno}
D.~G. Murray, M.~Isard, and Y.~Yu, ``{Steno: Automatic Optimization of
  Declarative Queries},'' in \emph{Proceedings of the 32nd ACM SIGPLAN
  Conference on Programming Language Design and Implementation (PLDI)}, 2011,
  pp. 121--131.

\bibitem{flumejava}
C.~Chambers, A.~Raniwala, F.~Perry, S.~Adams, R.~R. Henry, R.~Bradshaw, and
  N.~Weizenbaum, ``{{FlumeJava}: easy, efficient data-parallel pipelines},'' in
  \emph{ACM SIGPLAN Notices}, vol.~45, no.~6.\hskip 1em plus 0.5em minus
  0.4em\relax ACM, 2010, pp. 363--375.

\bibitem{musketeer}
I.~Gog, M.~Schwarzkopf, N.~Crooks, M.~P. Grosvenor, A.~Clement, and S.~Hand,
  ``{Musketeer: All for One, One for All in Data Processing Systems},'' in
  \emph{Proc. {ACM EuroSys}}, 2015, pp. 2:1--2:16.

\bibitem{emma}
A.~Alexandrov, A.~Kunft, A.~Katsifodimos, F.~Sch\"{u}ler, L.~Thamsen, O.~Kao,
  T.~Herb, and V.~Markl, ``{Implicit Parallelism Through Deep Language
  Embedding},'' in \emph{SIGMOD '15}, 2015.

\bibitem{tapir}
T.~B. Schardl, W.~S. Moses, and C.~E. Leiserson, ``{Tapir: Embedding Fork-Join
  Parallelism into LLVM's Intermediate Representation},'' in
  \emph{PPoPP}.\hskip 1em plus 0.5em minus 0.4em\relax ACM, 2017, pp. 249--265.

\bibitem{dphaskell}
\BIBentryALTinterwordspacing
{Data Parallel Haskell}. [Online]. Available:
  \url{https://wiki.haskell.org/GHC/Data\_Parallel\_Haskell}
\BIBentrySTDinterwordspacing

\bibitem{nesl}
G.~E. Blelloch, J.~C. Hardwick, S.~Chatterjee, J.~Sipelstein, and M.~Zagha,
  ``{Implementation of a Portable Nested Data-parallel Language},''
  \emph{SIGPLAN Not.}, vol.~28, no.~7, pp. 102--111, 1993.

\bibitem{delite}
H.~Chafi, A.~K. Sujeeth, K.~J. Brown, H.~Lee, A.~R. Atreya, and K.~Olukotun,
  ``{A Domain-specific Approach to Heterogeneous Parallelism},'' in
  \emph{Proceedings of the 16th ACM Symposium on Principles and Practice of
  Parallel Programming}, ser. PPoPP '11.\hskip 1em plus 0.5em minus 0.4em\relax
  New York, NY, USA: ACM, 2011, pp. 35--46.

\bibitem{delite-popl}
T.~Rompf, A.~K. Sujeeth, N.~Amin, K.~J. Brown, V.~Jovanovic, H.~Lee,
  M.~Jonnalagedda, K.~Olukotun, and M.~Odersky, ``{Optimizing Data Structures
  in High-level Programs: New Directions for Extensible Compilers Based on
  Staging},'' in \emph{POPL '13}, 2013.

\bibitem{dmll}
K.~J. Brown, H.~Lee, T.~Rompf, A.~K. Sujeeth, C.~De~Sa, C.~Aberger, and
  K.~Olukotun, ``{Have Abstraction and Eat Performance, Too: Optimized
  Heterogeneous Computing with Parallel Patterns},'' in \emph{Proceedings of
  the 2016 International Symposium on Code Generation and Optimization}, ser.
  CGO 2016.\hskip 1em plus 0.5em minus 0.4em\relax ACM, 2016, pp. 194--205.

\bibitem{cilk-reducers}
M.~Frigo, P.~Halpern, C.~E. Leiserson, and S.~Lewin-Berlin, ``{Reducers and
  Other Cilk++ Hyperobjects},'' in \emph{Proceedings of the Twenty-first Annual
  Symposium on Parallelism in Algorithms and Architectures}, ser. SPAA
  '09.\hskip 1em plus 0.5em minus 0.4em\relax New York, NY, USA: ACM, 2009, pp.
  79--90.

\bibitem{lvars}
L.~Kuper and R.~R. Newton, ``{LVars: Lattice-based Data Structures for
  Deterministic Parallelism},'' in \emph{Proceedings of the 2Nd ACM SIGPLAN
  Workshop on Functional High-performance Computing}, ser. FHPC '13.\hskip 1em
  plus 0.5em minus 0.4em\relax New York, NY, USA: ACM, 2013, pp. 71--84.

\end{thebibliography}


\end{document}